\documentclass[sigconf, screen]{acmart}

\usepackage{booktabs} 
\usepackage{multirow}
\usepackage{hyperref}

\setcopyright{acmlicensed}
\copyrightyear{2025}
\acmYear{2025}
\acmConference[MM '25]{{the 33th ACM International Conference on Multimedia}}{{October 27--31, 2025}}{{Dublin, Ireland}}

\begin{document}

\settopmatter{printacmref=false}
\renewcommand\footnotetextcopyrightpermission[1]{} 
\title{RTR-GS: 3D Gaussian Splatting for Inverse Rendering with Radiance Transfer and Reflection}

\author{Yongyang Zhou}
\email{yongyangzhou@bit.edu.cn}
\affiliation{%
  \institution{Beijing Institute of Technology}
  \city{Beijing}
  \country{China}
}

\author{Fang-Lue Zhang}
\email{fanglue.zhang@vuw.ac.nz}
\affiliation{%
  \institution{Victoria University of Wellington}
  \city{Wellington}
  \country{New Zealand}
}

\author{Zichen Wang}
\email{zichenwang@bit.edu.cn}
\affiliation{%
  \institution{Beijing Institute of Technology}
  \city{Beijing}
  \country{China}
}

\author{Lei Zhang}
\email{leizhang@bit.edu.cn}
\affiliation{%
  \institution{Beijing Institute of Technology}
  \city{Beijing}
  \country{China}
}






\begin{abstract}
3D Gaussian Splatting (3DGS) has demonstrated impressive capabilities in novel view synthesis. However, rendering reflective objects remains a significant challenge, particularly in inverse rendering and relighting. We introduce RTR-GS, a novel inverse rendering framework capable of robustly rendering objects with arbitrary reflectance properties, decomposing BRDF and lighting, and delivering credible relighting results. Given a collection of multi-view images, our method effectively recovers geometric structure through a hybrid rendering model that combines forward rendering for radiance transfer with deferred rendering for reflections. This approach successfully separates high-frequency and low-frequency appearances, mitigating floating artifacts caused by spherical harmonic overfitting when handling high-frequency details. We further refine BRDF and lighting decomposition using an additional physically-based deferred rendering branch. Experimental results show that our method enhances novel view synthesis, normal estimation, decomposition, and relighting while maintaining efficient training inference process. \href{https://github.com/ZyyZyy06/RTR-GS}{https://github.com/ZyyZyy06/RTR-GS}  
\end{abstract}




\begin{teaserfigure}
  \includegraphics[width=\textwidth]{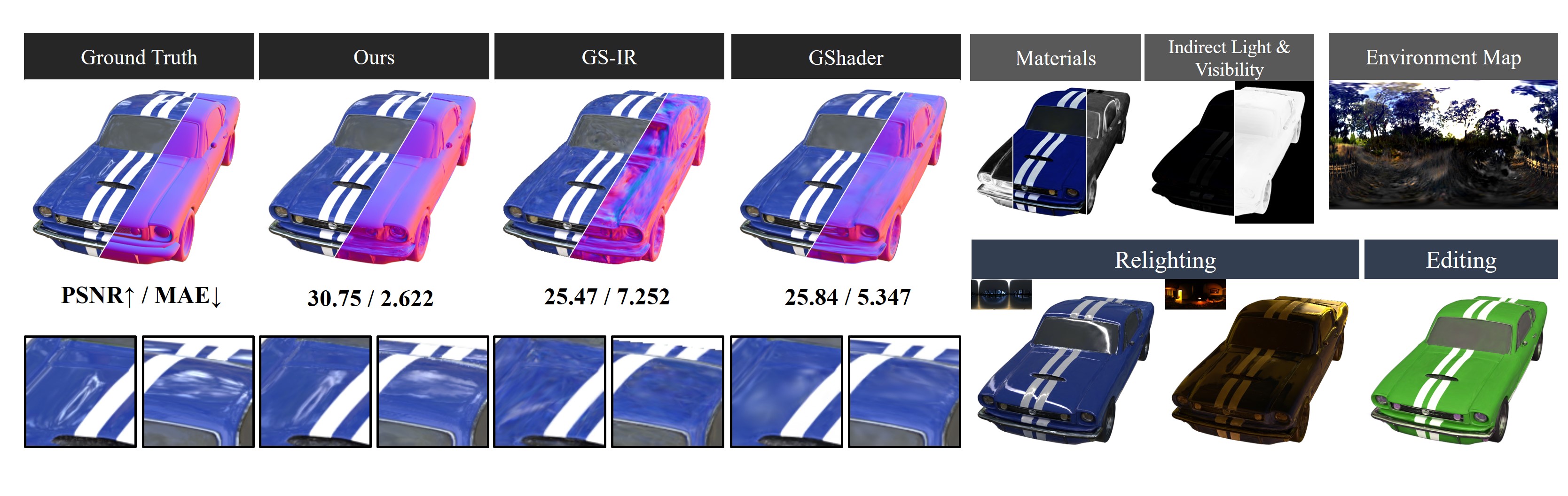}
  \caption{
  We propose RTR-GS, a framework for geometry-light-material decomposition from multi-view images. Our method significantly enhances normal estimation and visual realism for reflective surfaces compared to GS-IR \cite{gs-ir} and GShader \cite{gaussian-shader}. Additionally, we achieve material and lighting decomposition while accounting for secondary lighting effects through physically-based deferred rendering. The material components include albedo, metallic, and roughness. This high-quality decomposition enables realistic relighting and material editing.
  }
  \label{fig:teaser}
\end{teaserfigure}


\maketitle

\section{Introduction}
Inverse rendering is a long-standing challenge that seeks to decompose a 3D scene’s physical attributes—geometry, materials, and lighting—from captured images. This decomposition enables downstream tasks such as relighting and editing. The problem is particularly challenging due to the complex interplay of these attributes during rendering, especially under unknown illumination conditions, which make it inherently under-constrained.
Neural Radiance Fields (NeRF) \cite{nerf} have achieved remarkable success in novel view synthesis, laying the groundwork for inverse rendering. Methods such as \cite{nerfactor, physg, nerd, nero} use implicit neural representations, like Multi-Layer Perceptrons (MLPs), to decompose physical components. However, MLPs suffer from limited expressiveness and high computational costs, making it challenging to balance quality and efficiency.
3D Gaussian Splatting (3DGS) \cite{3dgs} improves both the speed and quality of learning-based volumetric rendering, and several methods \cite{gs-ir, r3dg, gir} have integrated physically-based rendering into this framework. However, spherical harmonic functions lack the directional resolution needed to accurately represent specular reflections, and overfitting during Gaussian splatting and cloning can introduce floating artifacts.

Accurate geometry is crucial for decomposing materials and lighting from complex appearances. However, high-frequency details can cause overfitting, leading to floating artifacts that deviate from physically smooth surfaces and compromise geometric accuracy.
To address this issue, we propose using a reflection map to store specular components, isolating high-frequency appearance details from the radiance component to mitigate overfitting. Additionally, we replace independent spherical harmonics with radiance transfer rendering, which imposes stronger global low-frequency constraints when computing radiance components. By separating high-frequency and low-frequency appearances, our method enables accurate recovery of geometric structures with arbitrary reflectance properties.
Following geometry reconstruction, we model occlusion and indirect illumination by baking visibility into 3D voxels and introducing indirect lighting parameters. This approach reduces aliasing artifacts in albedo, shadows, and lighting during decomposition. Finally, we achieve effective material and lighting decomposition by integrating an additional differentiable, physically-based deferred rendering branch.

The primary contribution of our work is the introduction of a Gaussian splatting-based inverse rendering framework, RTR-GS, which accurately estimates surface normals, bidirectional reflectance distribution functions (BRDF), and environmental lighting from multi-view images of both diffuse and specular objects. Specifically, it includes the following key aspects:

\begin{itemize} 
\item We propose a 3DGS-based hybrid rendering model that integrates reflection maps with radiance transfer, effectively separating high-frequency and low-frequency appearances. This enables efficient rendering of objects with arbitrary reflectance properties while reducing floating artifacts, thereby improving geometric structure recovery with high-quality normals.

\item We further enhance appearance decomposition through a dual-branch rendering approach, enabling efficient and accurate material and lighting decomposition via rational lighting modeling and occlusion data baked into 3D voxels.

\item Comprehensive experiments demonstrate that our method achieves state-of-the-art performance in novel view synthesis and relighting, producing credible results for both diffuse and specular objects.
\end{itemize}

\section{RELATED WORK}

\subsection{Neural representations}

Recent advancements in Neural Radiance Fields (NeRF) \cite{nerf} have garnered significant attention. Subsequent research has focused on enhancing rendering quality \cite{mip-nerf, zip-nerf, kou2025omniplane}, improving surface reconstruction \cite{volsdf, neus, neuralangelo}, and advancing object generation \cite{pi-gan, dreamfusion, prolific-dreamer, yuan2024munerf}, among other areas. Additionally, some methods aim to balance speed and quality \cite{instantNGP, k-planes, tensoRF, hexplane, dvgo, snerg}, facilitating more efficient evaluations.

3D Gaussian Splatting \cite{3dgs} effectively combines radiance field rendering with rasterization by leveraging discrete Gaussian distributions and the splatting technique. Subsequent research has focused on enhancing rendering quality \cite{scaffold-gs, mip-splatting}, more accurate geometry reconstruction \cite{gof, 2dgs, 3dgsr}, expanding editability \cite{luo20243d, zhang2024stylizedgs}, and increasing scalability \cite{nazarenus2024arbitrary}. However, these methods do not decompose appearance into materials and lighting, limiting their suitability for relighting and editing tasks. 

\begin{figure*}[htbp]
  \includegraphics[width=\textwidth]{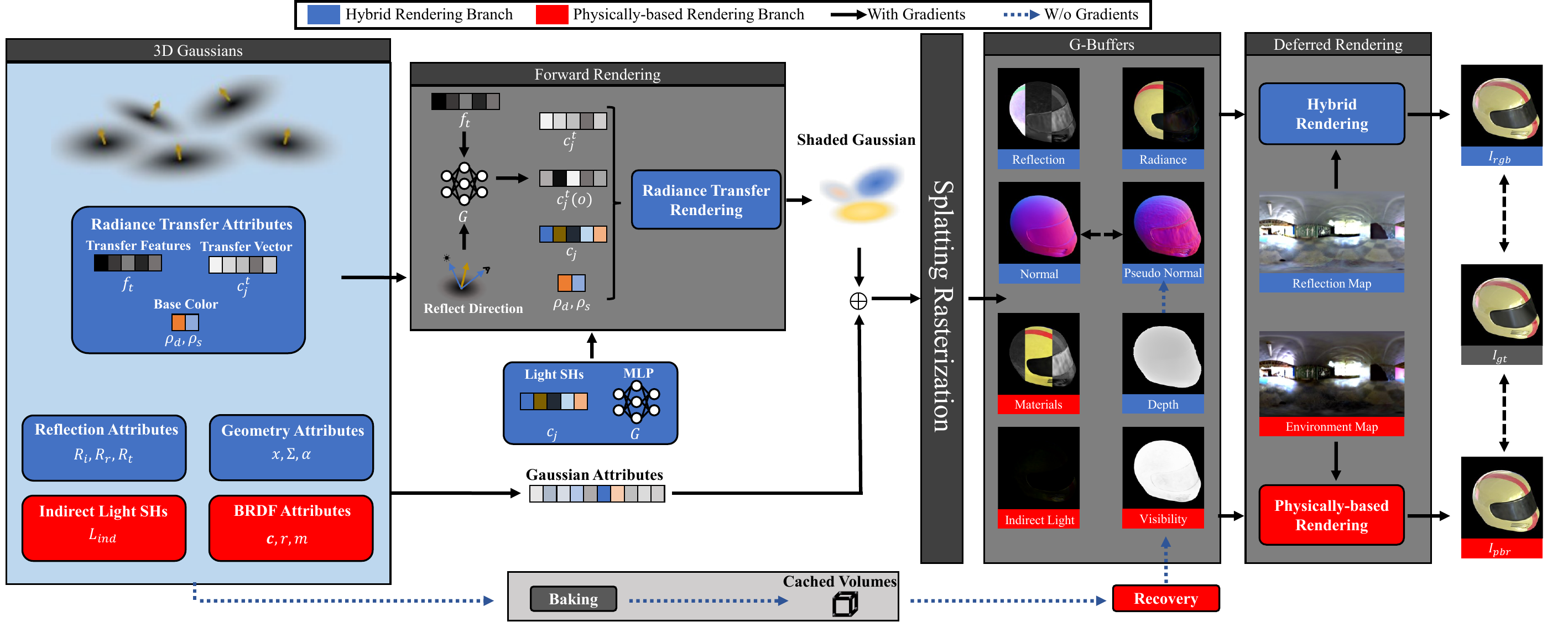}
  \caption{\textbf{RTR-GS Rendering Pipeline}.
  Our rendering pipeline consists of a hybrid rendering branch and a physically-based rendering branch. The hybrid rendering branch computes the radiance color for each Gaussian using forward rendering through radiance transfer, which is then blended with the reflections from deferred rendering after splatting. The physically-based rendering branch is fully implemented during the deferred rendering phase. Initially, the hybrid rendering branch reconstructs the fundamental geometric structure and stores visibility in voxel grids. The physically-based rendering branch is then activated to further decompose material appearances.
  }
  \label{fig:pipline}
\end{figure*}

\subsection{Inverse rendering}
Inverse rendering aims to decompose physically-based attributes from observations, including geometry, material, and lighting. A variety of methods simplify this problem by assuming controllable lighting conditions \cite{inv-1, inv-2, inv-3, inv-4, inv-5}. Some works relax these assumptions to consider direct lighting effects \cite{nerd, neural-pil, physg}. These works \cite{nerv, nerfactor, mii, neilf, neilf++, sire-ir} model secondary lighting effects using additional MLPs. To reduce computational overhead, some methods \cite{tensoir, tensosdf} employ tensor decomposition techniques inspired by TensoRF \cite{tensoRF}. For compatibility with existing rendering pipelines, NvDiffrec \cite{nvdiffrec} and NvDiffrecMC \cite{nvdiffrecMC} utilize differentiable rendering with rasterization or ray-tracing pipelines.

Methods based on 3D Gaussian Splatting (3DGS) have significantly accelerated training and rendering. GS-IR \cite{gs-ir}, GIR \cite{gir}, and R3DG \cite{r3dg} constrain surface normals using pseudo normals derived from depth and model shadows and indirect lighting through baking or ray-tracing. By leveraging pre-computed radiance transfer, PRT-GS \cite{prtgs} enables relighting, including secondary lighting effects. Phys3DGS \cite{phys3Dgs} integrates 3D Gaussian splats with mesh-based representations. Although these methods retain the high efficiency of 3DGS, using spherical harmonic functions as a radiance representation for geometry recovery often introduces floating artifacts on reflective surfaces, leading to geometric inaccuracies.

\subsection{Reflective object reconstruction}
Reconstructing reflective objects poses a significant challenge in inverse rendering tasks due to the high-frequency appearance changes that result in view inconsistencies. Ref-NeRF \cite{ref-nerf} tries to address this by using reflection directions instead of view directions and introducing Integrating Direction Encoding (IDE) to model reflections effectively. NeRO \cite{nero} explicitly models the reflection process. Spec-Gaussian \cite{spec-gaussian} simulates reflections using anisotropic Gaussians.
Deferred rendering approaches, such as DeferredGS \cite{deferredgs}, 3DGS-DR \cite{3dgs-dr}, GS-ROR \cite{gs-ror}, and GUS-IR\cite{gus-ir} replace forward rendering to better handle reflections. GaussianShader \cite{gaussian-shader} separates specular components and incorporates residual terms to capture secondary lighting effects. Additionally, PRD-GS \cite{PRD-GS} introduces progressive radiance distillation.

Inspired by these works, we adopt 3D Gaussians as the scene representation and develop an inverse rendering framework capable of effectively rendering object with arbitrary reflectance properties while also decomposing material and lighting components.

\section{Method}
\subsection{Overview}
\label{overview}
Figure \ref{fig:pipline} illustrates the overall framework of the proposed RTR-GS. We initialize 3D Gaussians using sparse point clouds generated randomly or estimated by COLMAP \cite{schonberger2016structure}. To model reflections, it is essential to define the normals for the Gaussians. We define normals as the shortest axis of each Gaussian, oriented toward the viewing direction, and optimize them synergistically using deferred rendering of reflections and pseudo-normals derived from a depth map (Sec. \ref{sec3.2}). Subsequently, we refine the Gaussians by introducing additional parameters and integrating key components into a hybrid rendering model (Sec. \ref{sec3.3}). This model combines radiance from forward rendering with reflections from deferred rendering, effectively separating high-frequency and low-frequency appearances to better represent complex materials and achieve high-quality scene reconstruction.
Next, we decompose the appearance using differentiable physically-based deferred rendering, incorporating occlusion baking, indirect lighting modeling, and additional BRDF parameters. During this process, we employ two rendering branches simultaneously to refine the geometry (Sec. \ref{sec 3.4}). Finally, we enhance the results through rendering losses and additional regularization terms (Sec. \ref{sec 3.5}).

\subsection{Deferred Rendering and Normal Modeling}
\label{sec3.2}

In the 3DGS framework, the attributes of multiple Gaussians are blended in the image plane using splatting and alpha blending, as follows:

\begin{equation}%
    \label{eqn:gsblend}
I_{f} = \sum_{i=0}^N f_{i}\alpha_iT_i
\end{equation}%

\noindent where $\alpha_i$ is the opacity, $T_i=\prod_{j=1}^{i-1}(1-\alpha_j)$ represents the accumulated transmittance, $f_i$ denotes the parameters of the $i$-th Gaussian, and $I_f$ represents the splatted screen-space attribute buffer. In vanilla 3DGS \cite{3dgs}, outgoing radiance is computed per-Gaussian before blending. This process is referred to as forward rendering. Additionally, the attributes associated with each Gaussian can be transformed into screen space for subsequent shading, a process known as deferred rendering. The following section explains our normal design and optimization based on the deferred rendering implementation.

Accurate normals are essential for modeling reflection. We define the normal direction as the shortest axis of the Gaussian. During the optimization process, the Gaussian shape typically flattens as it aligns with the surface, causing the shortest axis to correspond to a larger area. Similar to GS-IR \cite{gs-ir} and R3DG \cite{r3dg}, we optimize normals by enforcing consistency between the pseudo-normal map $\mathbf{\hat{n}_d}$, derived from the depth map, and the Gaussian normals map $\mathbf{n}$, as follows:

\begin{equation}%
    \label{eqn:normal_reg}
\mathcal{L}_n = \Vert \mathbf{n} - \mathbf{{\hat{n}_d}} \Vert_2
\end{equation}%

\begin{figure}[h!]
  \includegraphics[width=\linewidth]{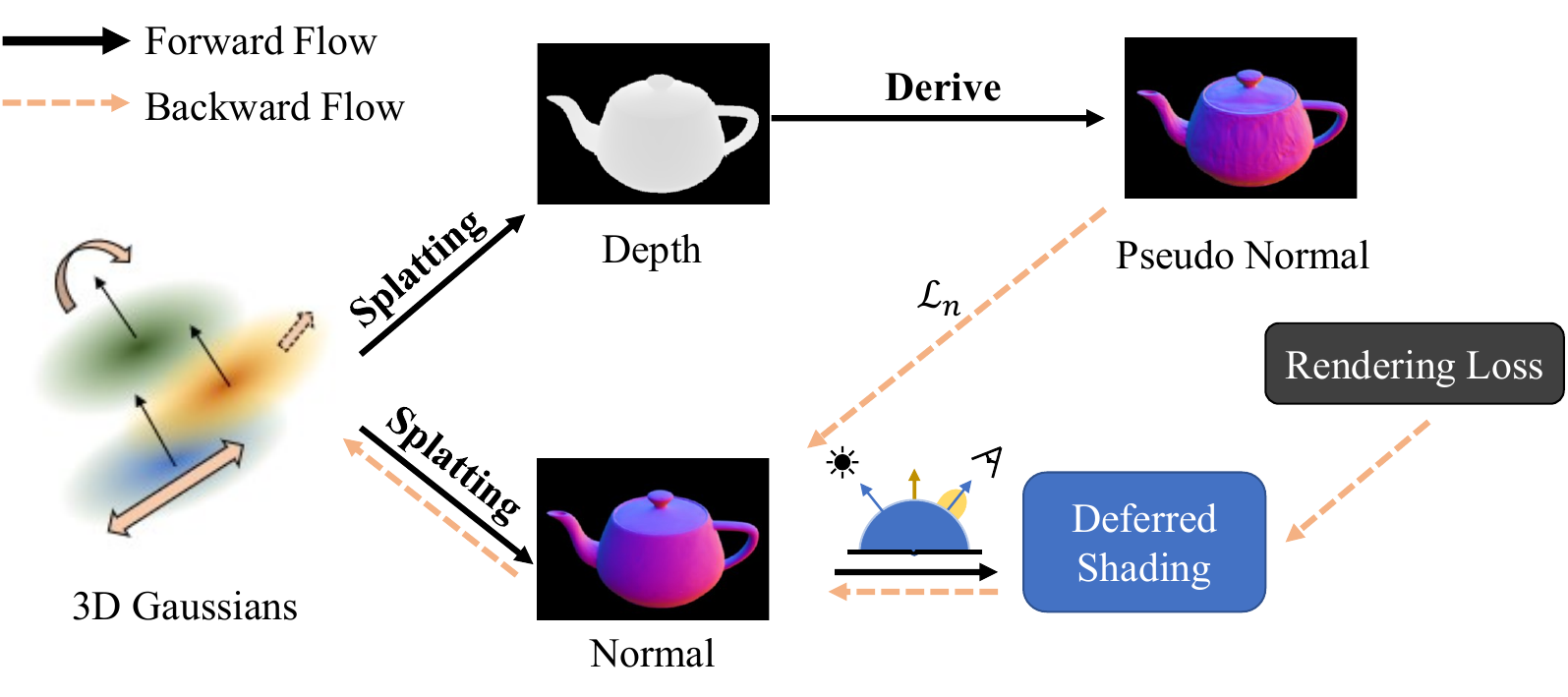}
  \caption{By adjusting the shapes of the Gaussians using the pseudo normals and gradients from the reflection map, the normals are optimized.}
  \label{fig:normal-adjust}
\end{figure}

\noindent This constraint is effective in optimizing normals when the depth map is smooth enough. Additionally, normals are used to compute reflection directions and contribute to deferred rendering. This process enables rendering losses to be backpropagated to the normals, refining the Gaussian shape.
When specular reflection is dominant, rendering losses from reflections primarily drive normal optimization. Conversely, in diffuse regions, depth-derived pseudo-normals impose a stronger constraint. Figure \ref{fig:normal-adjust} illustrates the normal optimization process.
Inspired by 3DGS-DR \cite{3dgs-dr}, we also introduce a simplified normal propagation mechanism that periodically enhances Gaussian opacity, improving the model’s robustness against extreme specular reflections.

\subsection{Hybrid Rendering and Radiance Transfer}
\label{sec3.3}
To effectively render appearances with diverse variations and to mitigate Gaussian floating artifacts caused by limited representation capability, we propose a hybrid rendering approach to replace the spherical harmonics-based forward rendering in 3DGS \cite{3dgs}. Our hybrid rendering model separates radiance and reflection to capture low-frequency and high-frequency components, respectively. Specifically, the radiance is computed using forward rendering, while the reflection is obtained through deferred rendering. The two components are then adaptively blended based on the reflection intensity as follows:

\begin{equation}%
    \label{eqn:radiance_blend}
I_{rgb} =  C_{r} \cdot (1.0 - {R_i}) + C_{ref} \cdot  {R_i}
\end{equation}%

\noindent where $C_r$ is the radiance color, $C_{ref}$ is the reflection color, and $R_i$ is the reflection intensity. The final blending is done in screen space. Further details on the reflection and radiance components are provided in the following sections.

\noindent \textbf{Reflection}. In forward rendering, BRDF lobes are computed individually using the respective normal of each Gaussian and are then blended after shading. However, this blending process broadens the final BRDF lobe, resulting in blurry rendering effects. In contrast, deferred rendering generates a single BRDF lobe based on the blended normal, providing higher precision and better preservation of BRDF sharpness. Similar observations have been analyzed in GUS-IR \cite{gus-ir} and GS-ROR \cite{gs-ror}.

For each Gaussian, we introduce additional reflection attributes for deferred rendering: reflection tint $R_{t}$ and reflection roughness $R_{r}$. We adopt a microfacet BRDF to simulate surfaces with varying roughness levels and achieve efficient computation using the split-sum approximation \cite{ue4}. The final reflection color is computed as:

\begin{equation}%
    \label{eqn:hr_reflection}
C_{ref} =  {R_t} \cdot {F_{ref}(E_r, {R_r}, \mathbf{n}, \mathbf{v})}
\end{equation}%

\noindent where $E_r$ is a learnable reflection map, $\mathbf{n}$ and $\mathbf{v}$ denote the normal and the view direction, respectively. $F_{ref}$ represents the split-sum approximation \cite{ue4}, which will be explained in more detail in Section \ref{sec 3.4}.

\noindent\textbf{Radiance}. Inspired by Precomputed Radiance Transfer (PRT) \cite{prt}, we adopt radiance transfer instead of spherical harmonics to compute outgoing radiance. Firstly, we will describe how radiance transfer is used to shade each Gaussian, including both view-independent and view-dependent components. Then we will explain the motivation behind using radiance transfer.

The view-independent component is consistent with the radiance transfer rendering in PRT. This calculation approximates the diffuse part of rendering equation as a dot product of two vectors as follows:
\begin{equation}%
    \label{eqn:radiance_transfer_diffuse}
{C_d} \approx \boldsymbol{\rho_d} \sum_{j=0}^{n^2}{c_j} c^t_j
\end{equation}%

\noindent where $\boldsymbol{\rho_d}$ represents the diffuse base color, $c_j$ denotes the coefficients of the spherical harmonics lighting, and $c^t_j$ represents the transfer vector. Notably, all Gaussians share the same spherical harmonics lighting $c_j$ but use individual transfer vector $c^t_j$.

For the view-dependent component, following the derivation in PRT \cite{prt}, we need to compute a radiance transfer matrix to convert environmental lighting into transferred lighting. However, $n$-order spherical harmonics lighting requires $n^2$ parameters to store the transfer matrix, leading to rapidly increasing storage costs as the number of Gaussians grows. To address this issue, we adopt neural radiance transfer for the view-dependent component and compute it in a manner similar to the view-independent case. Specifically, for each Gaussian, we introduce a set of randomly initialized radiance transfer features $f_t$ and a specular base color $\boldsymbol{\rho_s}$. We decode $f_t$ and the reflection direction $\mathbf{o}$ using a lightweight MLP $G$ to obtain the neural radiance transfer vector $c^t_j(\mathbf{o})$. The view-dependent outgoing radiance is computed as:

\begin{equation}%
    \label{eqn:radiance_transfer_specular}
{C_s(\mathbf{o})} \approx \boldsymbol{\rho_s}\sum_{j=0  }^{n^2}{c_j} c^t_j(\mathbf{o}), \quad with \quad c^t_j(\mathbf{o}) = G(f_t,\mathbf{o})
\end{equation}%

\noindent The total outgoing radiance is given by $C_r = C_d + C_s(\mathbf{o})$. After Gaussian splatting and blending, this radiance further participates in the blending process during deferred rendering. A detailed derivation of our radiance transfer implementation is provided in the supplementary materials.

\begin{figure}[h!]
  \includegraphics[width=0.9\linewidth]{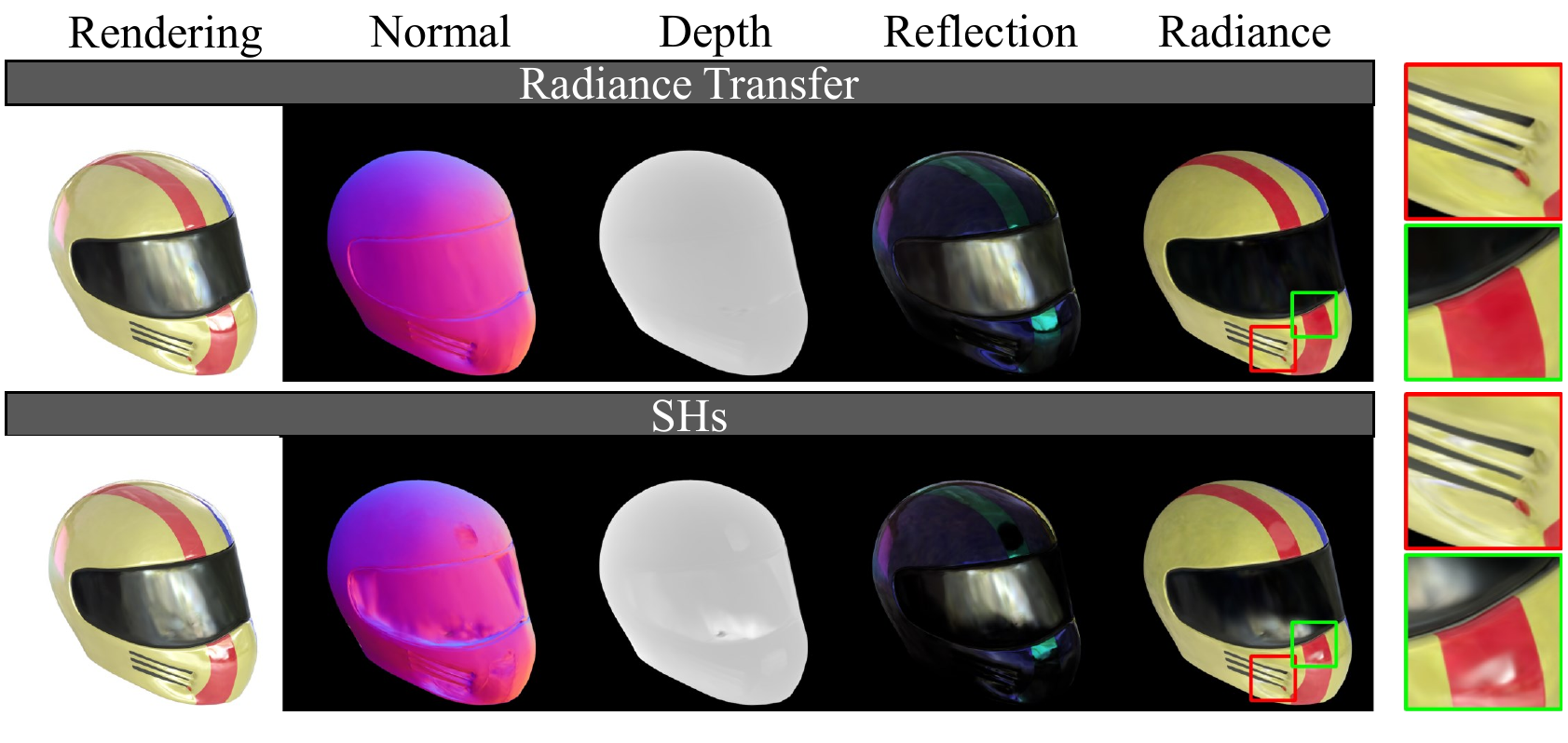}
  \caption{Radiance transfer provides a better representation of low-frequency appearances and helps prevent artifacts caused by overfitting high-frequency details.Such artifacts  can degrade the smoothness of depth and normal estimations, reducing the quality of the reconstructed geometry, and adversely affect subsequent decomposition processes.}
  \label{fig:radiance-transfer-sh}
\end{figure}

Compared to spherical harmonics, radiance transfer allows us to maintain enougth representational capacity while providing stronger global low-frequency constraints. In the shading process, all Gaussians share two global components: the spherical harmonics lighting $c_j$ and the MLP $G$. This design enables shading across Gaussians to be connected through shared components, promoting the representation of overall low-frequency variations. Meanwhile, each Gaussian has its own independent transfer vector and transfer features, along with base color attributes. This enables our radiance transfer representation to better handle components that are difficult to recover in the reflection part, such as local reflections and shadows. Figure \ref{fig:radiance-transfer-sh} illustrates the differences between our radiance transfer representation and spherical harmonics in modeling the radiance component. While the rendering results exhibit comparable visual quality, radiance transfer demonstrates better performance in low-frequency component fitting, prevents artifact generation, and maintains geometric smoothness.

\subsection{Illumination Modeling and Decomposition}
\label{sec 3.4}
We primarily use differentiable physically-based deferred rendering to decompose appearance into material and lighting components. To prevent aliasing artifacts in shadows, lighting, and albedo, we leverage the recovered geometric structure to bake occlusion information into a voxel grid, following the approach in GS-IR \cite{gs-ir}. Specifically, we set the background color to white and assign black to the Gaussian regions. The scene is then projected to generate a cubemap texture, which is converted into spherical harmonics coefficients and stored in the voxel grid. In the following, we describe our material and illumination modeling in detail.

For materials, we assign BRDF attributes to each Gaussian, including albedo $\mathbf{c}$, metallic $m$, and roughness $r$. For illumination, we use an environmental cubemap to implement image-based lighting (IBL) for handling direct lighting. Additionally, we add a parameter $L_{ind} \in [0,1]^3$ for each Gaussian to represent diffuse indirect lighting. The rendering equation $L(\mathbf{o}) = \int_{\Omega} L_i(\mathbf{i}) f(\mathbf{i}, \mathbf{o})(\mathbf{i} \cdot \mathbf{n})d{\mathbf{i}}$ is separated into diffuse and specular components to simplify computation. The diffuse component $L_d$ is computed as follows:

\begin{equation}%
    \label{eqn:pbr-diffuse}
    \begin{split}
    L_d (\textbf{x}) &= \frac{\boldsymbol{c}}{\pi}
    \int_{\Omega}{L_i(\mathbf{x},\mathbf{i}) (\mathbf{n} \cdot \mathbf{i}) d\mathbf{i}
    } \\
    &= \frac{\boldsymbol{c}}{\pi} 
    [\int_{\Omega}{L^{dir}_i(\mathbf{x}, \mathbf{i})(\mathbf{n} \cdot \mathbf{i}) d\mathbf{i}} + 
    \int_{\Omega} {L^{ind}_i(\mathbf{x}, \mathbf{i})(\mathbf{n}, \mathbf{i}) d\mathbf{i}}] \\
    &\approx \frac{\boldsymbol{c}}{\pi}
    [V(\mathbf{x}) L^{dir}_d(\mathbf{x})+ 
    (1-V(\mathbf{x})) L^{ind}_d(\mathbf{x}))]
    \end{split} 
\end{equation}%

\noindent where $L_d^{dir}(\mathbf{x})$ represents the direct environmental illumination, which depends only on the normal direction $\mathbf{n}$. This value is precomputed for efficiency and stored in a 2D texture. The indirect illumination $L_d^{ind}(\mathbf{x})$ is derived through the splatting and blending of $L_{ind}$. The visibility term $V(\mathbf{x})$ is determined by applying trilinear interpolation to the precomputed spherical harmonics stored in the baked voxel grid.

For the specular $L_s$, we employ the split-sum approximation \cite{ue4}, treating it as the product of two independent integrals as follows:

\begin{equation}%
    \label{eqn:pbr-specular}
    L_s(\textbf{x},\mathbf{o}) 
    \approx 
    \int_{\Omega} f_s(\mathbf{i},\mathbf{o})
    (\mathbf{n} \cdot \mathbf{i}) d\mathbf{i}
    \int_{\Omega}{ 
    L_i(\textbf{x}, \mathbf{i})D(\mathbf{i},\mathbf{o})(\mathbf{n} \cdot \mathbf{i}) d\mathbf{i}}
\end{equation}%

\noindent where $f(\mathbf{i}, \mathbf{o})$ represents the microfacet BRDF \cite{disney-brdf}. The first term of the integral represents the BRDF, which is independent of the lighting. It is precomputed and stored in a Look-Up Table (LUT). The second term accounts for the incoming radiance modulated by the normal distribution function (NDF) $D$, which is pre-integrated and represented using a filtered cubemap. 
Finally, the outgoing radiance is expressed as:

\begin{equation}%
    \label{eqn:pbr-final}
    L_o(\mathbf{x}, \mathbf{o}) = L_d(\mathbf{x})+ L_s(\mathbf{x}, \mathbf{o})
\end{equation}%

\noindent After completing deferred rendering, we obtain the final PBR result $I_{pbr}$.

In the decomposition process, we use both the previously mentioned hybrid rendering and PBR branches simultaneously, rather than freezing the geometric parameters or enabling only the PBR branch. This approach is adopted for two main reasons. Firstly, different rendering models still require corresponding geometric adjustments for proper adaptation, so completely freezing the geometric parameters is undesirable. We need to locally optimize the geometric attributes of the Gaussian to accommodate the newly introduced PBR branch. Secondly, since the PBR-related parameters are initialized randomly, using only PBR can easily lead to drastic changes in the geometric structure, which may render the baked visibility inapplicable. These two points will be further elaborated in the experimental section.

\subsection{Optimization}
\label{sec 3.5}
Throughout the training process, we optimize the geometric attributes of the Gaussian, as well as various rendering attributes closely related to the two rendering branches, as illustrated by the 3D Gaussians in Figure \ref{fig:pipline}. In addition, we need to optimize the small MLP $G$, which is a 3-layer network with 64 hidden units, used to decode the transfer feature and reflection direction, as well as two $6 \times 128 \times 128$ cubemaps: the reflection map for hybrid rendering and the environment map for PBR. We first activate the hybrid rendering branch and optimize the corresponding parameters. After restoring the basic geometric structure, we then activate the PBR branch and optimize all parameters. Finally, we outline the primary loss function and the specialized regularization terms.

\noindent\textbf{Rendering losses.} As in 3DGS\cite{3dgs}, we calculate the hybrid rendering loss $\mathcal{L}_{HR}$ and PBR loss $\mathcal{L}_{PBR}$ using the following equation:

\begin{equation}%
    \label{eqn:rendering_loss}
    \mathcal{L} = (1 - \lambda) \mathcal{L}_1(\hat{I},I_{gt}) + \lambda\mathcal{L}_{D-SSIM}(\hat{I}, I_{gt})
\end{equation}%

\noindent\textbf{Light regularization.} We apply a light regularization assuming a natural white incident light \cite{nero, nvdiffrec} for optimizing environment map used in PBR as follows:

\begin{equation}%
    \label{eqn:light_reg}
    \mathcal{L}_{light} = \sum_c(L_c - \frac{1}{3}
    \sum_c L_c), c \in \{R, G, B\}
\end{equation}%

\begin{table*}[htp!]
  \caption{NVS quality, training time and FPS on TensoIR, Shiny Blender and Stanford ORB datasets. “HR” represents our hybrid rendering branch.}
  \label{tab:nvs}
  \begin{tabular}{c|ccc|ccc|ccc|c|c}
  \hline
  \multirow{2}{*}{Methods} & \multicolumn{3}{c|}{TensoIR} & \multicolumn{3}{c|}{Shiny Blender} & \multicolumn{3}{c|}{Stanford ORB} & \multirow{2}{*}{Training Time}  & \multirow{2}{*}{FPS} \\
  & PSNR↑ & SSIM↑ & LPIPS↓ & PSNR↑ & SSIM↑ & LPIPS↓ & PSNR↑ & SSIM↑ & LPIPS↓ &\\
  \hline
  NeRO      & 32.60 & 0.933 & 0.082 & \underline{30.96} & 0.953 & \underline{0.081} & 29.25 & 0.970 & 0.060 & 8h &  <1\\
  TensoIR   & 35.18 & 0.976 & 0.040 & 27.95 & 0.896 & 0.159 & 34.81 & 0.983 & 0.029 & 5h & 4\\
  GS-IR     & 34.80 & 0.960 & 0.047 & 26.98 & 0.874 & 0.152 & 32.95 & 0.928 & 0.054 & 0.4h & 189\\
  R3DG      & 37.15 & 0.981 & 0.024 & 27.30 & 0.922 & 0.121 & 38.54 & 0.988 & 0.016 & 1h & 16\\
  3DGS-DR   & \underline{38.15} & 0.979 & 0.031 & 32.03 & \underline{0.960} & 0.084 & \underline{39.80} & 0.987 & \textbf{0.015} & 0.4h & 271\\
  GShader   & 37.13 & \underline{0.982} & \underline{0.023} & 30.87 & 0.953 & 0.088 & 36.02 & \underline{0.989} & 0.017 & 1h & 65\\
  Ours      & \textbf{39.17} & \textbf{0.985} & \textbf{0.021} & \textbf{33.99} & \textbf{0.971} & \textbf{0.061} & \textbf{39.81} & \textbf{0.990} & \underline{0.016} & 0.5h & 133\\
  \hline
  Ours(HR)  & 41.39 & 0.988 & 0.017 & 35.24 & 0.975 & 0.055 & 40.49 & 0.991 & 0.014 & 0.5h & 96\\
  
  \hline
  \end{tabular}
\end{table*}

\begin{table*}[htp!]
  \caption{Relighting quality is evaluated on the TensoIR, Shiny Blender, and Stanford ORB datasets.}
  \label{tab:relighting}
  \begin{tabular}{c|ccc|ccc|ccc}
  \hline
  \multirow{2}{*}{Methods} & \multicolumn{3}{c|}{TensoIR} & \multicolumn{3}{c|}{Shiny Blender} & \multicolumn{3}{c}{Stanford ORB} \\
  & PSNR↑ & SSIM↑ & LPIPS↓ & PSNR↑ & SSIM↑ & LPIPS↓ & PSNR↑ & SSIM↑ & LPIPS↓ \\
  \hline
    TensoIR   & \underline{28.55} & \textbf{0.945} & 0.080 & \underline{22.30} & 0.842 & 0.184 & 26.22 & 0.947 & 0.049 \\
    GShader   & 26.86 & 0.930 & 0.063 & 19.20 & 0.874 & \underline{0.131} & 26.23 & 0.952 & 0.043 \\
    GS-IR     & 25.98 & 0.897 & 0.092 & 21.18 & 0.846 & 0.160 & \underline{28.44} & \underline{0.960} & \underline{0.038} \\
    R3DG      & 28.52 & 0.931 & \underline{0.069} & 20.69 & \underline{0.869} & 0.141 & 27.88 & 0.957 & 0.039 \\
    Ours & \textbf{30.10} & \underline{0.944} & \textbf{0.053} & \textbf{26.16} & \textbf{0.928} & \textbf{0.084} & \textbf{28.93} & \textbf{0.967} & \textbf{0.029} \\
  \hline
  \end{tabular}
\end{table*}

\begin{figure*}[tbp]
    \includegraphics[width=0.85\textwidth]{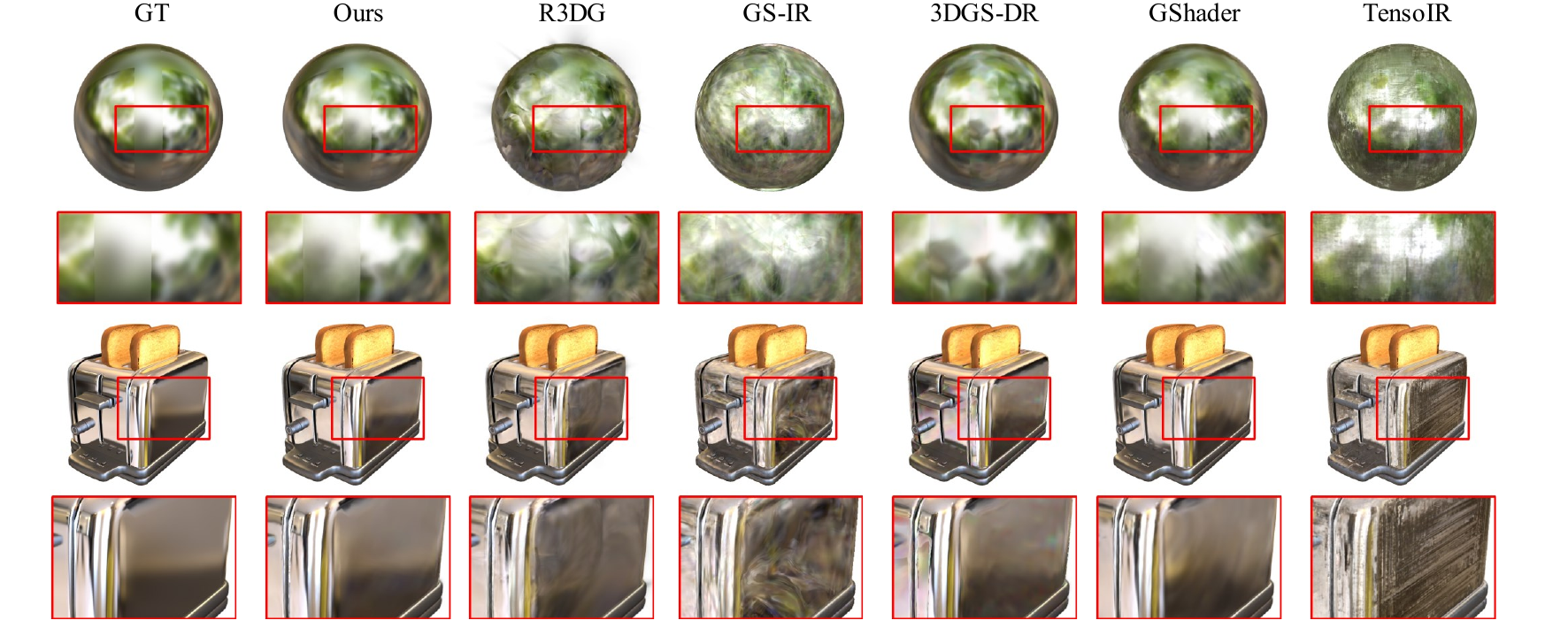}
  \caption{Qualitative comparisons on a synthetic dataset. Our method retains more details, particularly in specular regions.}
  \label{fig:nvs-compare}
\end{figure*}

\noindent \textbf{Metal reflection prior.}
Due to the reflective properties of metals, we aim to make the metallic parameter $m$ in the PBR model as close as possible to the reflection intensity $R_i$ in hybrid rendering, as follows:

\begin{equation}%
    \label{eqn:metal_reflection_prior}
    \mathcal{L}_{m} = \mathcal{L}_1(m, R_i)
\end{equation}%

\noindent which encourages our two rendering branches to maintain appearance consistency in high-frequency regions. The effectiveness of this regularization term is discussed in the following section. In addition, we incorporate a bilateral smoothness term $\mathcal{L}_{s}$ and an object mask constraint $\mathcal{L}_{o}$. The final loss $\mathcal{L}$ is defined as:

\begin{equation}%
    \label{eqn:fianl_loss}
    \mathcal{L} = \mathcal{L}_{HR} + \lambda_{PBR}\mathcal{L}_{PBR} + \lambda_0 \mathcal{L}_{light} + \lambda_1 \mathcal{L}_{m} + \lambda_2 \mathcal{L}_{n} + \mathcal{L}_{s} + \mathcal{L}_{o}
\end{equation}%

\noindent where $\lambda_{PBR} =0$ or $1$, $\lambda_0 = 0.003$, $\lambda_1 = 0.1$, $\lambda_2=0.02$. Detailed descriptions of $\mathcal{L}_{s}$ and $\mathcal{L}_{o}$ are provided in the supplementary materials. 

\section{Experiments}

\subsection{Evaluation Setup}

\textbf{Dataset and Metrics.} For synthetic objects in the TensoIR \cite{tensoir} and Shiny Blender \cite{ref-nerf} datasets, as well as real objects in the Stanford ORB dataset \cite{stanfordORB}, we evaluate the performance of novel view synthesis and relighting using PSNR, SSIM \cite{ssim}, and LPIPS \cite{lpips} metrics. For the ball object in the Shiny Blender dataset, only qualitative results are provided due to the absence of relighting ground truth (GT). In addition, we use mean angular error (MAE) to evaluate the quality of normal estimation. In addition, we have also provided the results of training duration and inference speed (FPS). We further evaluate novel view synthesis on the Ref-Real \cite{ref-nerf} and MipNeRF-360 \cite{mip-nerf360} datasets. \textbf{Numbers} in bold represent the best performance, while \underline{underscored} numbers indicate the second-best performance.

\textbf{Methods for Comparison.}
We compared the quality of novel view synthesis against several NeRF-based methods \cite{nero, tensoir} and 3DGS-based methods \cite{gs-ir, r3dg, gaussian-shader, 3dgs-dr}. In addition, we evaluated the relighting quality between different inverse rendering methods. All methods were implemented and trained using their publicly available code and default configurations.

\begin{figure*}[tbp]
\includegraphics[width=0.85\textwidth]{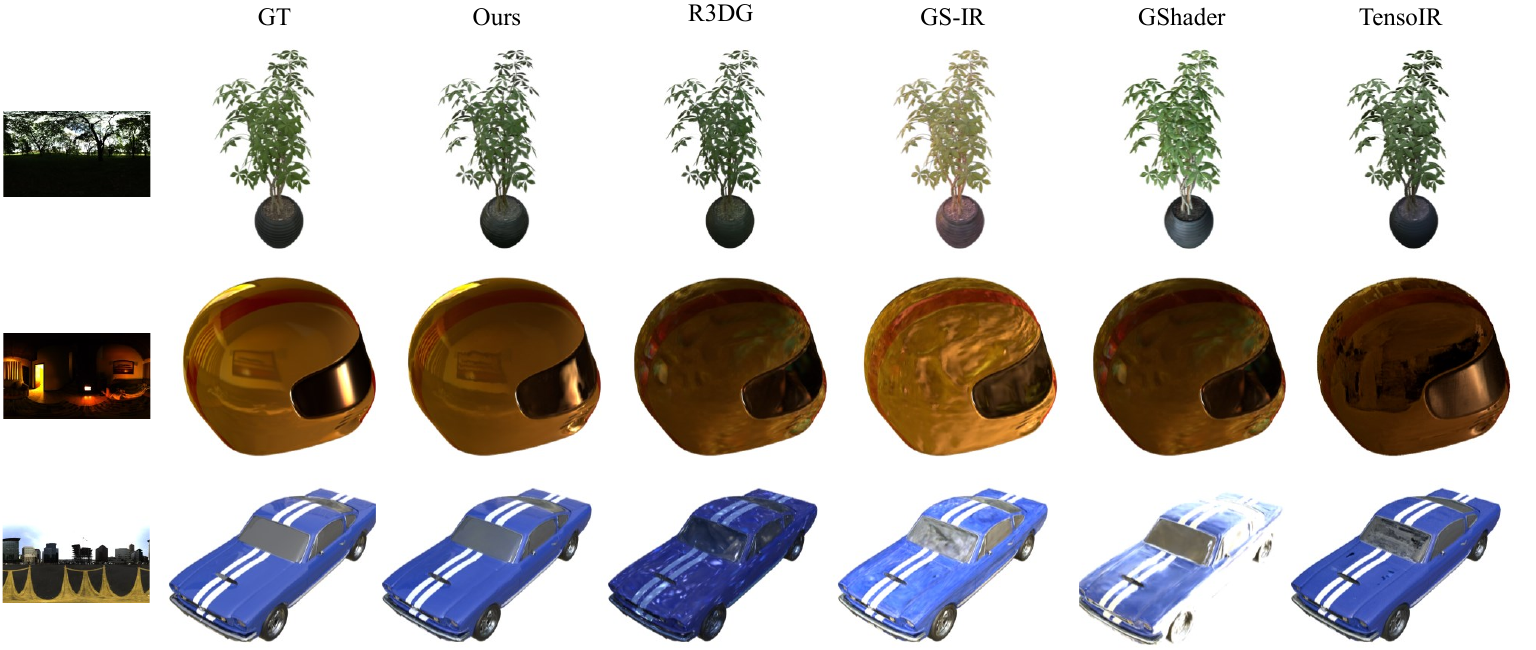}
  \caption{Qualitative comparisons of relighting with different environment lighting conditions.}
  \label{fig:relgiht-compare}
\end{figure*}

\begin{figure*}[tbp]
  \includegraphics[width=0.85\textwidth]{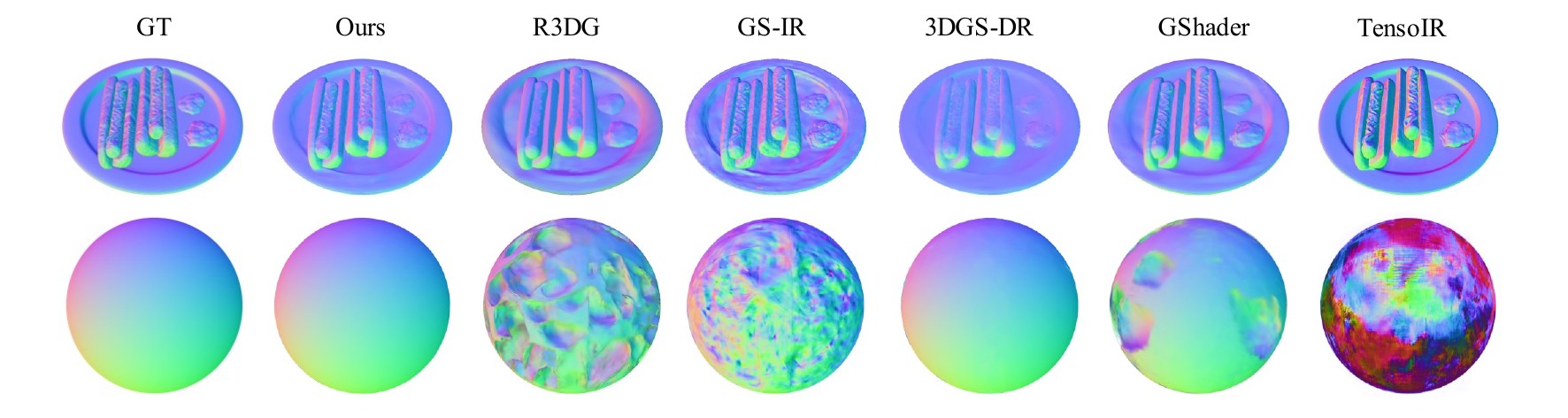}
  \caption{Qualitative comparisons of normal produced by different methods. Our method provides robust normal estimation.
  }
  \label{fig:normale-compare}
\end{figure*}

\subsection{Comparison with previous works}

\textbf{Novel view synthesis.}
Table \ref{tab:nvs} presents the quantitative comparison results for novel view synthesis (NVS) on object-level datasets. Our PBR results show clear advantages over other methods. Additionally, we provide our Hybrid Rendering (HR) branch results to demonstrate the effectiveness of the hybrid rendering model. Visual comparisons are provided in Figure \ref{fig:nvs-compare}. Notably, our method preserves stable geometric structures even with high-frequency surface variations, producing clearer and more accurate novel views. Furthermore, Table \ref{tab: nvs real scene} presents our results on the Ref-Real dataset \cite{ref-nerf} and the Mip-NeRF 360 dataset \cite{mip-nerf360}, where our method achieves competitive quantitative results.

\textbf{Relighting.}
Table \ref{tab:relighting} presents the results of the relighting comparison. For the TensoIR and Shiny Blender datasets, albedo is aligned to the ground truth via channel-wise scaling before relighting as described in \cite{physg, stanfordORB}. For the Stanford ORB dataset, albedo scaling is disabled to more accurately evaluate absolute decomposition performance on real objects. Results for the TensoIR and Shiny Blender datasets are averaged over all viewpoints under five different environment maps. For the Stanford ORB dataset, relighting is evaluated using the provided 20 image-environment map pairs. Visual comparisons are provided in  Figure \ref{fig:relgiht-compare}. Our method's superior detail preservation and effectively suppresses aliasing artifacts in both albedo and lighting, leading to more realistic and visually consistent relighting results. Notably, our approach maintains credibility under different relighting conditions, without significant surface artifacts appearing on either rough or smooth objects.

\begin{figure}[tbp]
\includegraphics[width=\linewidth]{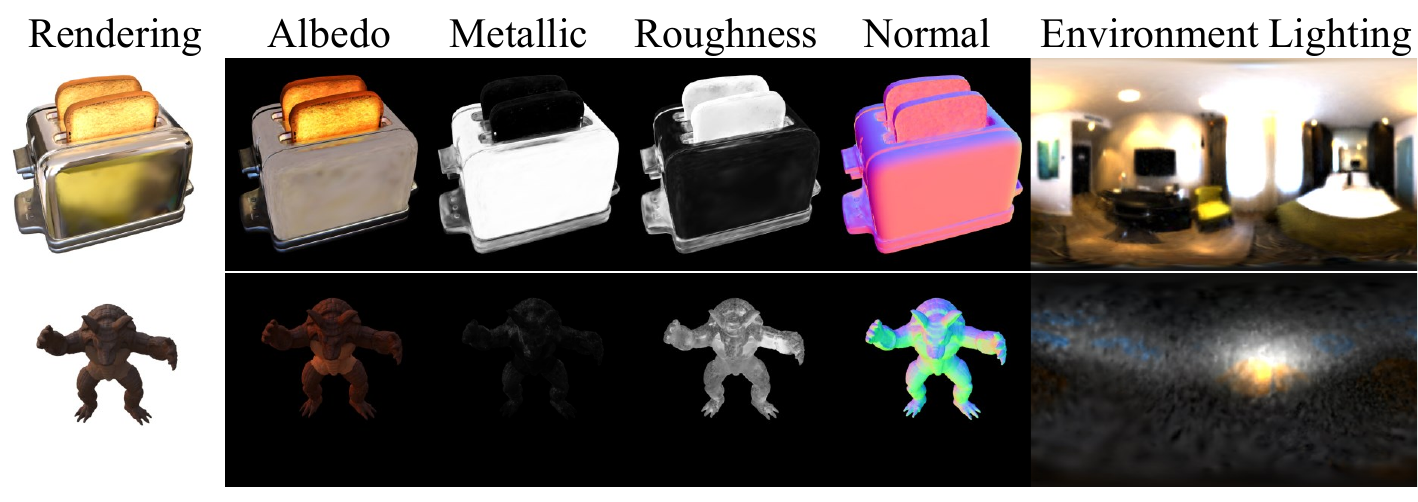}
  \caption{Normal, albedo, roughness, metallic and environment lighing results on synthetic dataset.}
  \label{fig:decomposition}
\end{figure}

\textbf{Normal and materials estimation.}
Table \ref{tab: MAE} and Figure \ref{fig:normale-compare} present the results of our normal estimation. Notably, in the presence of high-frequency surface details, our method effectively prevents surface discontinuities caused by floating artifacts. In Figure \ref{fig:decomposition}, we visualize the estimated albedo, metallic, roughness, normal, and environmental lighting components. Our framework successfully decomposes both diffuse and specular objects. For specular objects, we achieve high-quality decomposition results with clearer environmental lighting. Additional albedo estimation results and more qualitative comparisons are provided in the supplementary materials.

\begin{table}[h!]
  \caption{ Novel view synthesis quality evaluated using PSNR, SSIM, and LPIPS on the Ref-Real dataset and the Mip-NeRF 360 dataset.}
  \label{tab: nvs real scene}
  \begin{tabular}{c|ccc|ccc}
  \hline
  \multirow{2}{*}{Methods} &  \multicolumn{3}{c|}{Ref-Real} & \multicolumn{3}{c}{Mip-NeRF 360} \\
    & PSNR↑ & SSIM↑ & LPIPS↓ & PSNR↑ & SSIM↑ & LPIPS↓ \\
  \hline
   GS-IR   & 23.41 & 0.606 & \textbf{0.297} & \underline{26.18} & \underline{0.801} & \textbf{0.200}  \\
   GShader   & 21.13 & 0.578 & 0.375 & 22.33 & 0.577 & 0.329 \\
   3DGS-DR   & \underline{23.51} & \textbf{0.638} & 0.343 & 25.14 & 0.783 & 0.304\\
   Ours   & \textbf{23.54} & \underline{0.627} & \underline{0.337} & \textbf{26.65} & \textbf{0.806} & \underline{0.233}\\
  \hline
  \end{tabular}
\end{table}

\begin{table}[h!]
  \caption{Normal estimation quality with Gaussian-based methods evaluated using MAE↓ on the TensoIR dataset and the Shiny Blender dataset.}
  \label{tab: MAE}
  \begin{tabular}{c|cccccc}
  \hline
    & GS-IR & R3DG & 3DGS-DR & GShader & Ours \\
  \hline
  TensoIR & \underline{5.313} & 5.914 & 5.728 & \textbf{5.303} & 5.347\\
  Shiny Blender  & 9.328 & 9.238 & \underline{3.632} & 4.800 & \textbf{3.091} \\
  \hline
  \end{tabular}
\end{table}

\begin{table}[h!]
  \caption{Ablation study of key components on the Shiny Blender dataset. "w/o radiance transfer" represents using SHs to calculate the radiance part in hybrid rendering. "Propagation" denotes simplified normal propagation. "Frozen geometry" indicates freezing geometry attributes during decomposition. "w/o hybrid rendering" refers to disabling the hybrid rendering branch during decomposition.}
  \label{tab:ablations-shinyblender}
  \begin{tabular}{c|c|c}
  \hline
  {Ablations} & NVS PSNR↑ & Relighting PSNR↑ \\
  \hline
  ours                      & \textbf{33.99} & \textbf{26.16} \\
  w/o radiance transfer     & 32.15 & 25.85 \\
  w/o propagation           & 33.26 & 26.09 \\
  w/o $\mathcal{L}_{m}$     & 33.76 & 25.88 \\
  \hline
  w/ frozen geometry      & 31.49 & 24.66 \\
  w/o hybrid rendering      & 32.90 & 25.18 \\
  \hline
  \end{tabular}
\end{table}

\begin{figure}[h!]
  \includegraphics[width=\linewidth]{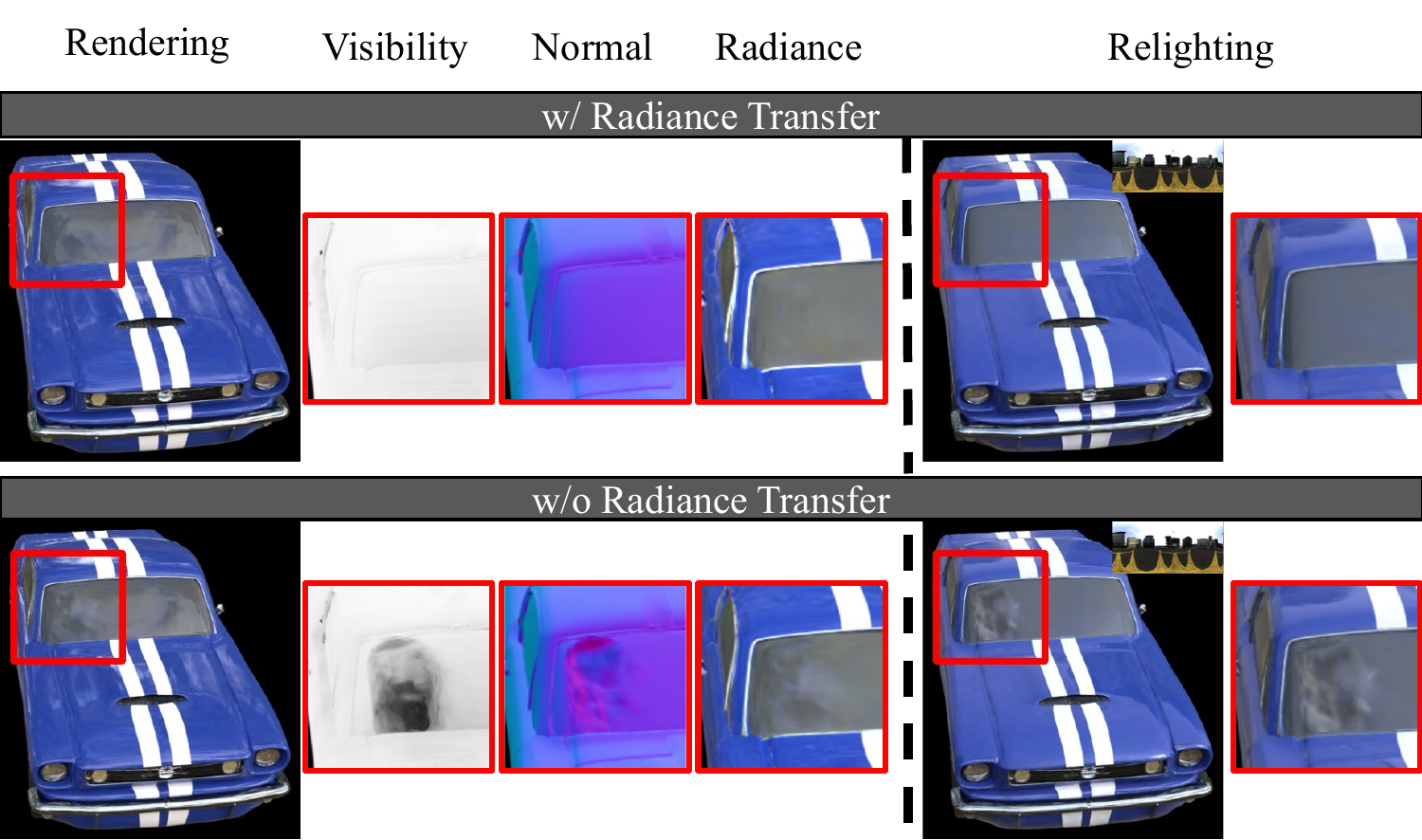}
  \caption{Radiance transfer can more effectively separate low-frequency components of appearance, thereby preventing artifacts caused by overfitting. These artifacts compromise geometric smoothness and degrade the quality of rendering and relighting.}
  \label{fig:ablation_radiance_transfer}
\end{figure}

\subsection{Ablation Study}
We specifically evaluated the effectiveness of radiance transfer compared to spherical harmonics. Additionally, we performed ablation studies on simplified normal propagation to validate the contribution of our proposed components. We also evaluate the impact of the metal reflection prior introduced in Sec.~\ref{sec 3.5}. For decomposition process, we further conducted experiments of using fixed geometric parameters and disabling the hybrid rendering branch (i.e., using only the PBR branch) during appearance decomposition, to demonstrate the advantages of our dual-branch rendering framework.

\textbf{{Analysis on radiance transfer.}}
As illustrated in Figure \ref{fig:ablation_radiance_transfer}, using radiance transfer instead of spherical harmonics to represent the radiance component in hybrid rendering reduces floating artifacts and prevents normal and visibility errors caused by local geometric inaccuracies, particularly for specular objects. These improvements significantly enhance the quality of relighting. As shown in Table \ref{tab:ablations-shinyblender}, radiance transfer also leads to notable improvements in quantitative results.

\textbf{{Analysis on decomposition process.}}
When decomposing the appearance, we simultaneously enable hybrid rendering and PBR to fine-tune the geometry, making it compatible with both rendering models. We also evaluate the effects of freezing geometric parameters or enabling only the PBR branch, which demonstrates the limitations of single-branch approaches. As shown in Table \ref{tab:ablations-shinyblender}, both frozen geometry and enabling the PBR branch only lead to significant quality degradation. The former occurs because the geometric structure required for hybrid rendering does not fully meet PBR's requirements, while the latter leads to geometric mutations, rendering the baked occlusion ineffective.

\begin{figure}[h!]
  \includegraphics[width=0.95\linewidth]{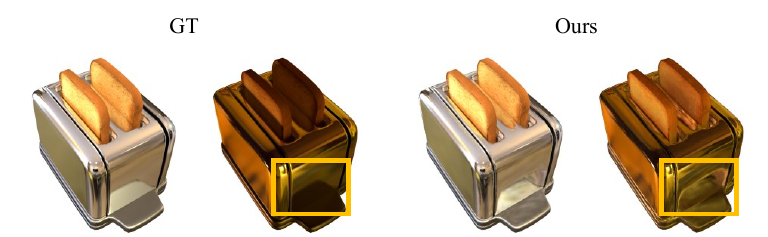}
  \caption{Limitation of our method.}
  \label{fig:limitation}
\end{figure}

\textbf{Limitation}
We assume that lighting originates from an infinite distance, which differs from actual lighting conditions in large-scale scenes. Additionally, our method does not consider more complex indirect lighting effects, such as inter-reflections. These limitations are shown in Figure \ref{fig:limitation}.

\section{Conclusions}
We introduce RTR-GS, an inverse rendering framework that enables realistic novel view synthesis and relighting through Gaussian splatting and deferred rendering. By separating high-frequency and low-frequency appearances using reflection maps and radiance transfer, we achieve high-quality hybrid rendering and normal estimation. Building on this, we further decompose material and lighting from the appearance by an additional PBR branch. Experimental results demonstrate that our method delivers competitive performance in novel view synthesis and relighting across various objects.
In the future, we aim to explore more precise rendering techniques and incorporate more complex secondary lighting effects.






\bibliographystyle{ACM-Reference-Format}
\bibliography{sample-base}

@String{Computer = "{IEEE} Computer" }

@String{Springer = "Springer-Verlag" }

@inproceedings{schonberger2016structure,
  title={Structure-from-motion revisited},
  author={Schonberger, Johannes L and Frahm, Jan-Michael},
  booktitle={Proceedings of the IEEE conference on computer vision and pattern recognition},
  pages={4104--4113},
  year={2016}
}

@inproceedings{3dgs-dr,
  title={3d gaussian splatting with deferred reflection},
  author={Ye, Keyang and Hou, Qiming and Zhou, Kun},
  booktitle={ACM SIGGRAPH 2024 Conference Papers},
  pages={1--10},
  year={2024}
}

@article{nerfactor,
  title={Nerfactor: Neural factorization of shape and reflectance under an unknown illumination},
  author={Zhang, Xiuming and Srinivasan, Pratul P and Deng, Boyang and Debevec, Paul and Freeman, William T and Barron, Jonathan T},
  journal={ACM Transactions on Graphics (ToG)},
  volume={40},
  number={6},
  pages={1--18},
  year={2021},
  publisher={ACM New York, NY, USA}
}

@inproceedings{nerd,
  title={Nerd: Neural reflectance decomposition from image collections},
  author={Boss, Mark and Braun, Raphael and Jampani, Varun and Barron, Jonathan T and Liu, Ce and Lensch, Hendrik},
  booktitle={Proceedings of the IEEE/CVF International Conference on Computer Vision},
  pages={12684--12694},
  year={2021}
}

@article{nero,
  title={Nero: Neural geometry and brdf reconstruction of reflective objects from multiview images},
  author={Liu, Yuan and Wang, Peng and Lin, Cheng and Long, Xiaoxiao and Wang, Jiepeng and Liu, Lingjie and Komura, Taku and Wang, Wenping},
  journal={ACM Transactions on Graphics (TOG)},
  volume={42},
  number={4},
  pages={1--22},
  year={2023},
  publisher={ACM New York, NY, USA}
}

@inproceedings{gs-ir,
  title={Gs-ir: 3d gaussian splatting for inverse rendering},
  author={Liang, Zhihao and Zhang, Qi and Feng, Ying and Shan, Ying and Jia, Kui},
  booktitle={Proceedings of the IEEE/CVF Conference on Computer Vision and Pattern Recognition},
  pages={21644--21653},
  year={2024}
}

@inproceedings{r3dg,
  title={Relightable 3D Gaussians: realistic point cloud relighting with BRDF decomposition and ray tracing},
  author={Gao, Jian and Gu, Chun and Lin, Youtian and Li, Zhihao and Zhu, Hao and Cao, Xun and Zhang, Li and Yao, Yao},
  booktitle={European Conference on Computer Vision},
  pages={73--89},
  year={2025},
  organization={Springer}
}

@article{gir,
  title={Gir: 3d gaussian inverse rendering for relightable scene factorization},
  author={Shi, Yahao and Wu, Yanmin and Wu, Chenming and Liu, Xing and Zhao, Chen and Feng, Haocheng and Liu, Jingtuo and Zhang, Liangjun and Zhang, Jian and Zhou, Bin and others},
  journal={arXiv preprint arXiv:2312.05133},
  year={2023}
}

@inproceedings{gaussian-shader,
  title={Gaussianshader: 3d gaussian splatting with shading functions for reflective surfaces},
  author={Jiang, Yingwenqi and Tu, Jiadong and Liu, Yuan and Gao, Xifeng and Long, Xiaoxiao and Wang, Wenping and Ma, Yuexin},
  booktitle={Proceedings of the IEEE/CVF Conference on Computer Vision and Pattern Recognition},
  pages={5322--5332},
  year={2024}
}

@article{nerf,
  title={Nerf: Representing scenes as neural radiance fields for view synthesis},
  author={Mildenhall, Ben and Srinivasan, Pratul P and Tancik, Matthew and Barron, Jonathan T and Ramamoorthi, Ravi and Ng, Ren},
  journal={Communications of the ACM},
  volume={65},
  number={1},
  pages={99--106},
  year={2021},
  publisher={ACM New York, NY, USA}
}

@article{volsdf,
  title={Volume rendering of neural implicit surfaces},
  author={Yariv, Lior and Gu, Jiatao and Kasten, Yoni and Lipman, Yaron},
  journal={Advances in Neural Information Processing Systems},
  volume={34},
  pages={4805--4815},
  year={2021}
}

@article{neus,
  title={Neus: Learning neural implicit surfaces by volume rendering for multi-view reconstruction},
  author={Wang, Peng and Liu, Lingjie and Liu, Yuan and Theobalt, Christian and Komura, Taku and Wang, Wenping},
  journal={arXiv preprint arXiv:2106.10689},
  year={2021}
}

@inproceedings{neuralangelo,
  title={Neuralangelo: High-fidelity neural surface reconstruction},
  author={Li, Zhaoshuo and M{\"u}ller, Thomas and Evans, Alex and Taylor, Russell H and Unberath, Mathias and Liu, Ming-Yu and Lin, Chen-Hsuan},
  booktitle={Proceedings of the IEEE/CVF Conference on Computer Vision and Pattern Recognition},
  pages={8456--8465},
  year={2023}
}

@inproceedings{mip-nerf,
  title={Mip-nerf: A multiscale representation for anti-aliasing neural radiance fields},
  author={Barron, Jonathan T and Mildenhall, Ben and Tancik, Matthew and Hedman, Peter and Martin-Brualla, Ricardo and Srinivasan, Pratul P},
  booktitle={Proceedings of the IEEE/CVF International Conference on Computer Vision},
  pages={5855--5864},
  year={2021}
}

@inproceedings{zip-nerf,
  title={Zip-nerf: Anti-aliased grid-based neural radiance fields},
  author={Barron, Jonathan T and Mildenhall, Ben and Verbin, Dor and Srinivasan, Pratul P and Hedman, Peter},
  booktitle={Proceedings of the IEEE/CVF International Conference on Computer Vision},
  pages={19697--19705},
  year={2023}
}

@inproceedings{pi-gan,
  title={pi-gan: Periodic implicit generative adversarial networks for 3d-aware image synthesis},
  author={Chan, Eric R and Monteiro, Marco and Kellnhofer, Petr and Wu, Jiajun and Wetzstein, Gordon},
  booktitle={Proceedings of the IEEE/CVF Conference on Computer Vision and Pattern Recognition},
  pages={5799--5809},
  year={2021}
}

@article{dreamfusion,
  title={Dreamfusion: Text-to-3d using 2d diffusion},
  author={Poole, Ben and Jain, Ajay and Barron, Jonathan T and Mildenhall, Ben},
  journal={arXiv preprint arXiv:2209.14988},
  year={2022}
}

@article{prolific-dreamer,
  title={Prolificdreamer: High-fidelity and diverse text-to-3d generation with variational score distillation},
  author={Wang, Zhengyi and Lu, Cheng and Wang, Yikai and Bao, Fan and Li, Chongxuan and Su, Hang and Zhu, Jun},
  journal={Advances in Neural Information Processing Systems},
  volume={36},
  year={2024}
}

@inproceedings{k-planes,
  title={K-planes: Explicit radiance fields in space, time, and appearance},
  author={Fridovich-Keil, Sara and Meanti, Giacomo and Warburg, Frederik Rahb{\ae}k and Recht, Benjamin and Kanazawa, Angjoo},
  booktitle={Proceedings of the IEEE/CVF Conference on Computer Vision and Pattern Recognition},
  pages={12479--12488},
  year={2023}
}

@inproceedings{tensoRF,
  title={Tensorf: Tensorial radiance fields},
  author={Chen, Anpei and Xu, Zexiang and Geiger, Andreas and Yu, Jingyi and Su, Hao},
  booktitle={European Conference on Computer Vision},
  pages={333--350},
  year={2022},
  organization={Springer}
}

@inproceedings{hexplane,
  title={Hexplane: A fast representation for dynamic scenes},
  author={Cao, Ang and Johnson, Justin},
  booktitle={Proceedings of the IEEE/CVF Conference on Computer Vision and Pattern Recognition},
  pages={130--141},
  year={2023}
}

@article{deferredgs,
  title={DeferredGS: Decoupled and editable Gaussian splatting with deferred shading},
  author={Wu, Tong and Sun, Jiamu and Lai, Yukun and Ma, Yuewen and Kobbelt, Leif and Gao, Lin},
  journal={arXiv preprint arXiv:2404.09412},
  year={2024}
}

@article{instantNGP,
  title={Instant neural graphics primitives with a multiresolution hash encoding},
  author={M{\"u}ller, Thomas and Evans, Alex and Schied, Christoph and Keller, Alexander},
  journal={ACM Transactions on Graphics (TOG)},
  volume={41},
  number={4},
  pages={1--15},
  year={2022},
  publisher={ACM New York, NY, USA}
}

@article{kou2025omniplane,
  title={OmniPlane: A recolorable representation for dynamic scenes in omnidirectional videos},
  author={Kou, Simin and Zhang, Fang-Lue and Nazarenus, Jakob and Koch, Reinhard and Dodgson, Neil A},
  journal={IEEE Transactions on Visualization and Computer Graphics},
  year={2025},
  publisher={IEEE}
}

@article{yuan2024munerf,
  title={Munerf: Robust makeup transfer in neural radiance fields},
  author={Yuan, Yu-Jie and Han, Xinyang and He, Yue and Zhang, Fang-Lue and Gao, Lin},
  journal={IEEE Transactions on Visualization and Computer Graphics},
  year={2024},
  publisher={IEEE}
}

@inproceedings{dvgo,
  title={Direct voxel grid optimization: Super-fast convergence for radiance fields reconstruction},
  author={Sun, Cheng and Sun, Min and Chen, Hwann-Tzong},
  booktitle={Proceedings of the IEEE/CVF Conference on Computer Vision and Pattern Recognition},
  pages={5459--5469},
  year={2022}
}

@inproceedings{snerg,
  title={Baking neural radiance fields for real-time view synthesis},
  author={Hedman, Peter and Srinivasan, Pratul P and Mildenhall, Ben and Barron, Jonathan T and Debevec, Paul},
  booktitle={Proceedings of the IEEE/CVF International Conference on Computer Vision},
  pages={5875--5884},
  year={2021}
}

@article{3dgs,
  title={3D Gaussian splatting for real-time radiance field rendering.},
  author={Kerbl, Bernhard and Kopanas, Georgios and Leimk{\"u}hler, Thomas and Drettakis, George},
  journal={ACM Trans. Graph.},
  volume={42},
  number={4},
  pages={139--1},
  year={2023}
}

@article{spec-gaussian,
  title={Spec-gaussian: Anisotropic view-dependent appearance for 3d gaussian splatting},
  author={Yang, Ziyi and Gao, Xinyu and Sun, Yangtian and Huang, Yihua and Lyu, Xiaoyang and Zhou, Wen and Jiao, Shaohui and Qi, Xiaojuan and Jin, Xiaogang},
  journal={arXiv preprint arXiv:2402.15870},
  year={2024}
}

@article{3dgsr,
  title={3dgsr: Implicit surface reconstruction with 3d gaussian splatting},
  author={Lyu, Xiaoyang and Sun, Yang-Tian and Huang, Yi-Hua and Wu, Xiuzhe and Yang, Ziyi and Chen, Yilun and Pang, Jiangmiao and Qi, Xiaojuan},
  journal={arXiv preprint arXiv:2404.00409},
  year={2024}
}

@article{gof,
  title={Gaussian opacity fields: Efficient and compact surface reconstruction in unbounded scenes},
  author={Yu, Zehao and Sattler, Torsten and Geiger, Andreas},
  journal={arXiv preprint arXiv:2404.10772},
  year={2024}
}

@inproceedings{2dgs,
  title={2d Gaussian splatting for geometrically accurate radiance fields},
  author={Huang, Binbin and Yu, Zehao and Chen, Anpei and Geiger, Andreas and Gao, Shenghua},
  booktitle={ACM SIGGRAPH 2024 Conference Papers},
  pages={1--11},
  year={2024}
}

@article{nazarenus2024arbitrary,
  title={Arbitrary optics for Gaussian splatting using space warping},
  author={Nazarenus, Jakob and Kou, Simin and Zhang, Fang-Lue and Koch, Reinhard},
  journal={Journal of Imaging},
  volume={10},
  number={12},
  pages={330},
  year={2024},
  publisher={MDPI}
}

@inproceedings{luo20243d,
  title={3D Gaussian editing with a single image},
  author={Luo, Guan and Xu, Tian-Xing and Liu, Ying-Tian and Fan, Xiao-Xiong and Zhang, Fang-Lue and Zhang, Song-Hai},
  booktitle={Proceedings of the 32nd ACM International Conference on Multimedia},
  pages={6627--6636},
  year={2024}
}

@article{zhang2024stylizedgs,
  title={Stylizedgs: Controllable stylization for 3d gaussian splatting},
  author={Zhang, Dingxi and Yuan, Yu-Jie and Chen, Zhuoxun and Zhang, Fang-Lue and He, Zhenliang and Shan, Shiguang and Gao, Lin},
  journal={arXiv preprint arXiv:2404.05220},
  year={2024}
}

@inproceedings{scaffold-gs,
  title={Scaffold-gs: Structured 3d gaussians for view-adaptive rendering},
  author={Lu, Tao and Yu, Mulin and Xu, Linning and Xiangli, Yuanbo and Wang, Limin and Lin, Dahua and Dai, Bo},
  booktitle={Proceedings of the IEEE/CVF Conference on Computer Vision and Pattern Recognition},
  pages={20654--20664},
  year={2024}
}

@inproceedings{inv-1,
  title={Inverse path tracing for joint material and lighting estimation},
  author={Azinovic, Dejan and Li, Tzu-Mao and Kaplanyan, Anton and Nie{\ss}ner, Matthias},
  booktitle={Proceedings of the IEEE/CVF Conference on Computer Vision and Pattern Recognition},
  pages={2447--2456},
  year={2019}
}

@inproceedings{inv-2,
  title={Deep reflectance volumes: Relightable reconstructions from multi-view photometric images},
  author={Bi, Sai and Xu, Zexiang and Sunkavalli, Kalyan and Ha{\v{s}}an, Milo{\v{s}} and Hold-Geoffroy, Yannick and Kriegman, David and Ramamoorthi, Ravi},
  booktitle={Computer Vision--ECCV 2020: 16th European Conference, Glasgow, UK, August 23--28, 2020, Proceedings, Part III 16},
  pages={294--311},
  year={2020},
  organization={Springer}
}

@inproceedings{inv-3,
  title={Deep 3d capture: Geometry and reflectance from sparse multi-view images},
  author={Bi, Sai and Xu, Zexiang and Sunkavalli, Kalyan and Kriegman, David and Ramamoorthi, Ravi},
  booktitle={Proceedings of the IEEE/CVF Conference on Computer Vision and Pattern Recognition},
  pages={5960--5969},
  year={2020}
}

@article{inv-4,
  title={The relightables: Volumetric performance capture of humans with realistic relighting},
  author={Guo, Kaiwen and Lincoln, Peter and Davidson, Philip and Busch, Jay and Yu, Xueming and Whalen, Matt and Harvey, Geoff and Orts-Escolano, Sergio and Pandey, Rohit and Dourgarian, Jason and others},
  journal={ACM Transactions on Graphics (ToG)},
  volume={38},
  number={6},
  pages={1--19},
  year={2019},
  publisher={ACM New York, NY, USA}
}

@inproceedings{inv-5,
  title={On joint estimation of pose, geometry and svbrdf from a handheld scanner},
  author={Schmitt, Carolin and Donne, Simon and Riegler, Gernot and Koltun, Vladlen and Geiger, Andreas},
  booktitle={Proceedings of the IEEE/CVF Conference on Computer Vision and Pattern Recognition},
  pages={3493--3503},
  year={2020}
}

@article{nerv,
  title={Nerv: Neural representations for videos},
  author={Chen, Hao and He, Bo and Wang, Hanyu and Ren, Yixuan and Lim, Ser Nam and Shrivastava, Abhinav},
  journal={Advances in Neural Information Processing Systems},
  volume={34},
  pages={21557--21568},
  year={2021}
}

@article{neural-pil,
  title={Neural-pil: Neural pre-integrated lighting for reflectance decomposition},
  author={Boss, Mark and Jampani, Varun and Braun, Raphael and Liu, Ce and Barron, Jonathan and Lensch, Hendrik},
  journal={Advances in Neural Information Processing Systems},
  volume={34},
  pages={10691--10704},
  year={2021}
}

@inproceedings{physg,
  title={Physg: Inverse rendering with spherical gaussians for physics-based material editing and relighting},
  author={Zhang, Kai and Luan, Fujun and Wang, Qianqian and Bala, Kavita and Snavely, Noah},
  booktitle={Proceedings of the IEEE/CVF Conference on Computer Vision and Pattern Recognition},
  pages={5453--5462},
  year={2021}
}

@inproceedings{mii,
  title={Modeling indirect illumination for inverse rendering},
  author={Zhang, Yuanqing and Sun, Jiaming and He, Xingyi and Fu, Huan and Jia, Rongfei and Zhou, Xiaowei},
  booktitle={Proceedings of the IEEE/CVF Conference on Computer Vision and Pattern Recognition},
  pages={18643--18652},
  year={2022}
}

@inproceedings{neilf,
  title={Neilf: Neural incident light field for physically-based material estimation},
  author={Yao, Yao and Zhang, Jingyang and Liu, Jingbo and Qu, Yihang and Fang, Tian and McKinnon, David and Tsin, Yanghai and Quan, Long},
  booktitle={European Conference on Computer Vision},
  pages={700--716},
  year={2022},
  organization={Springer}
}

@inproceedings{neilf++,
  title={Neilf++: Inter-reflectable light fields for geometry and material estimation},
  author={Zhang, Jingyang and Yao, Yao and Li, Shiwei and Liu, Jingbo and Fang, Tian and McKinnon, David and Tsin, Yanghai and Quan, Long},
  booktitle={Proceedings of the IEEE/CVF International Conference on Computer Vision},
  pages={3601--3610},
  year={2023}
}

@article{sire-ir,
  title={Sire-ir: Inverse rendering for brdf reconstruction with shadow and illumination removal in high-illuminance scenes},
  author={Yang, Ziyi and Chen, Yanzhen and Gao, Xinyu and Yuan, Yazhen and Wu, Yu and Zhou, Xiaowei and Jin, Xiaogang},
  journal={arXiv preprint arXiv:2310.13030},
  year={2023}
}

@inproceedings{tensoir,
  title={Tensoir: Tensorial inverse rendering},
  author={Jin, Haian and Liu, Isabella and Xu, Peijia and Zhang, Xiaoshuai and Han, Songfang and Bi, Sai and Zhou, Xiaowei and Xu, Zexiang and Su, Hao},
  booktitle={Proceedings of the IEEE/CVF Conference on Computer Vision and Pattern Recognition},
  pages={165--174},
  year={2023}
}

@article{tensosdf,
  title={Tensosdf: Roughness-aware tensorial representation for robust geometry and material reconstruction},
  author={Li, Jia and Wang, Lu and Zhang, Lei and Wang, Beibei},
  journal={ACM Transactions on Graphics (TOG)},
  volume={43},
  number={4},
  pages={1--13},
  year={2024},
  publisher={ACM New York, NY, USA}
}

@inproceedings{nvdiffrec,
  title={Extracting triangular 3d models, materials, and lighting from images},
  author={Munkberg, Jacob and Hasselgren, Jon and Shen, Tianchang and Gao, Jun and Chen, Wenzheng and Evans, Alex and M{\"u}ller, Thomas and Fidler, Sanja},
  booktitle={Proceedings of the IEEE/CVF Conference on Computer Vision and Pattern Recognition},
  pages={8280--8290},
  year={2022}
}

@article{nvdiffrecMC,
  title={Shape, light, and material decomposition from images using monte carlo rendering and denoising},
  author={Hasselgren, Jon and Hofmann, Nikolai and Munkberg, Jacob},
  journal={Advances in Neural Information Processing Systems},
  volume={35},
  pages={22856--22869},
  year={2022}
}

@inproceedings{prtgs,
  title={PRTGS: Precomputed radiance transfer of gaussian Splats for real-time high-quality relighting},
  author={Guo, Yijia and Bai, Yuanxi and Hu, Liwen and Guo, Ziyi and Liu, Mianzhi and Cai, Yu and Huang, Tiejun and Ma, Lei},
  booktitle={Proceedings of the 32nd ACM International Conference on Multimedia},
  pages={5112--5120},
  year={2024}
}

@article{phys3Dgs,
  title={Phys3DGS: Physically-based 3D Gaussian splatting for inverse rendering},
  author={Choi, Euntae and Yoo, Sungjoo},
  journal={arXiv preprint arXiv:2409.10335},
  year={2024}
}

@inproceedings{ref-nerf,
  title={Ref-nerf: Structured view-dependent appearance for neural radiance fields},
  author={Verbin, Dor and Hedman, Peter and Mildenhall, Ben and Zickler, Todd and Barron, Jonathan T and Srinivasan, Pratul P},
  booktitle={2022 IEEE/CVF Conference on Computer Vision and Pattern Recognition (CVPR)},
  pages={5481--5490},
  year={2022},
  organization={IEEE}
}

@article{gs-ror,
  title={Gs-ror: 3d gaussian splatting for reflective object relighting via sdf priors},
  author={Zhu, Zuo-Liang and Wang, Beibei and Yang, Jian},
  journal={arXiv preprint arXiv:2406.18544},
  year={2024}
}

@article{PRD-GS,
  title={Progressive radiance distillation for inverse rendering with Gaussian splatting},
  author={Ye, Keyang and Hou, Qiming and Zhou, Kun},
  journal={arXiv preprint arXiv:2408.07595},
  year={2024}
}

@article{ue4,
  title={Real shading in unreal engine 4},
  author={Karis, Brian and Games, Epic},
  journal={Proc. Physically Based Shading Theory Practice},
  volume={4},
  number={3},
  pages={1},
  year={2013}
}

@incollection{prt,
  title={Precomputed radiance transfer for real-time rendering in dynamic, low-frequency lighting environments},
  author={Sloan, Peter-Pike and Kautz, Jan and Snyder, John},
  booktitle={Seminal Graphics Papers: Pushing the Boundaries, Volume 2},
  pages={339--348},
  year={2023}
}

@article{ssim,
  title={Image quality assessment: from error visibility to structural similarity},
  author={Wang, Zhou and Bovik, Alan C and Sheikh, Hamid R and Simoncelli, Eero P},
  journal={IEEE Transactions on Image Processing},
  volume={13},
  number={4},
  pages={600--612},
  year={2004},
  publisher={IEEE}
}

@inproceedings{lpips,
  title={The unreasonable effectiveness of deep features as a perceptual metric},
  author={Zhang, Richard and Isola, Phillip and Efros, Alexei A and Shechtman, Eli and Wang, Oliver},
  booktitle={Proceedings of the IEEE Conference on Computer Vision and Pattern Recognition},
  pages={586--595},
  year={2018}
}

@inproceedings{mip-nerf360,
  title={Mip-nerf 360: Unbounded anti-aliased neural radiance fields},
  author={Barron, Jonathan T and Mildenhall, Ben and Verbin, Dor and Srinivasan, Pratul P and Hedman, Peter},
  booktitle={Proceedings of the IEEE/CVF Conference on Computer Vision and Pattern Recognition},
  pages={5470--5479},
  year={2022}
}

@inproceedings{disney-brdf,
  title={Physically-based shading at disney},
  author={Burley, Brent and Studios, Walt Disney Animation},
  booktitle={Acm Siggraph},
  volume={2012},
  pages={1--7},
  year={2012},
  organization={vol. 2012}
}

@inproceedings{mip-splatting,
  title={Mip-splatting: Alias-free 3d gaussian splatting},
  author={Yu, Zehao and Chen, Anpei and Huang, Binbin and Sattler, Torsten and Geiger, Andreas},
  booktitle={Proceedings of the IEEE/CVF Conference on Computer Vision and Pattern Recognition},
  pages={19447--19456},
  year={2024}
}

@article{gus-ir,
  title={GUS-IR: Gaussian splatting with unified shading for inverse rendering},
  author={Liang, Zhihao and Li, Hongdong and Jia, Kui and Guo, Kailing and Zhang, Qi},
  journal={arXiv preprint arXiv:2411.07478},
  year={2024}
}

@article{stanfordORB,
  title={Stanford-orb: a real-world 3d object inverse rendering benchmark},
  author={Kuang, Zhengfei and Zhang, Yunzhi and Yu, Hong-Xing and Agarwala, Samir and Wu, Elliott and Wu, Jiajun and others},
  journal={Advances in Neural Information Processing Systems},
  volume={36},
  pages={46938--46957},
  year={2023}
}










\appendix
\newpage
\section{radiance transfer}

\subsection{View-independent part}
We will derive the calculation formula for radiance transfer starting from the rendering equation as follows:

\begin{equation}%
    \label{eqn:render-equation}
{L(\mathbf{o})} = \int{L(\mathbf{\boldsymbol{\omega}})f(\boldsymbol{\omega},\mathbf{o})}\mathrm{max}(0,\mathbf{n} \cdot \boldsymbol{\omega})d\mathbf{\boldsymbol{\omega}} 
\end{equation}%

\noindent We transform the integration domain from the upper hemisphere to the entire sphere by constraining the cosine value. Assuming that spherical harmonics are used to reconstruct the incident radiance $L(\mathbf{\boldsymbol{\omega}})$, the radiance transfer term $T(\boldsymbol{\omega},\mathbf{o})$ accounts for the remaining components. This transformation allows for a more comprehensive representation of the radiance transfer and incident light interaction. The rendering equation can be approximated in the following form:

\begin{equation}%
    \label{eqn:view_independent_1}
{L(\mathbf{o})} \approx \sum_{j=0}^n{c_j} \int{B_j(\boldsymbol{\omega})} T(\boldsymbol{\omega}, \mathbf{o})d\boldsymbol{\omega} 
\end{equation}%

\noindent where $B_j(\boldsymbol{\omega})$ is the corresponding basis function. For the view-independent part, the radiance transfer term is decomposed into the diffuse albedo $\boldsymbol{\rho_d}$ and $T'(i)$, as follows:

\begin{equation}%
    \label{eqn:view_independent_2}
{L_d} \approx \boldsymbol{\rho_d} \sum_{j=0}^{n^2}{c_j} \int{B_j(\boldsymbol{\omega})} T'(\boldsymbol{\omega})d\mathbf{\boldsymbol{\omega}} 
\end{equation}%

\noindent where the integral part can be considered as a projection of $T'(\boldsymbol{\omega})$ on the basis functions. Through projection, the corresponding spherical harmonic coefficients can be calculated as follows:

\begin{equation}%
    \label{eqn:view_independent_3}
    c^t_j =  \int{B_j(\boldsymbol{\omega})} T'(\boldsymbol{\omega})d\mathbf{\boldsymbol{\omega}}
\end{equation}%

In the case of using a finite order of spherical harmonic functions, the outgoing radiance can be approximated as:

\begin{equation}%
    \label{eqn:view_independent_4}
{L_d} \approx \boldsymbol{\rho_d} \sum_{j=0}^{n^2}{c_j} c^t_j
\end{equation}%

By using basis functions as an intermediate representation, we can quickly approximate the complex integral through a point multiplication of two sets of coefficient vectors.

\subsection{View-dependent part}
For the view-dependent part, similar to Equation (\ref{eqn:view_independent_2}), the radiance transfer term is simultaneously related to both the incident and outgoing directions.
\begin{equation}%
    \label{eqn:view_dependent_1}
{L_s(\mathbf{o})} \approx \boldsymbol{\rho}_s\sum_{j=1}^{n^2}{c_j} \int{B_j(\boldsymbol{\omega})} T'(\boldsymbol{\omega},\mathbf{o})d\mathbf{\boldsymbol{\omega}} 
\end{equation}%

Similar to Equation (\ref{eqn:view_independent_3}), we project the radiance transfer term onto the spherical harmonics and obtain the corresponding coefficients as Eqution (\ref{eqn:view_dependent_2}). 

\begin{equation}%
    \label{eqn:view_dependent_2}
    c^t_j(\mathbf{o}) =  \int{B_j(\boldsymbol{\omega})} T'(\boldsymbol{\omega},\mathbf{o})d\mathbf{\boldsymbol{\omega}}
\end{equation}%

However, the coefficients in this case are related to the outgoing direction and can not be used directly as follows:

\begin{equation}%
    \label{eqn:view_dependent_3}
{L_s}(\mathbf{o}) \approx \boldsymbol{\rho}_s\sum_{j=1}^{n^2}{c_j} c^t_j(\mathbf{o})
\end{equation}%

Therefore, we continue to project $c^t_j(\mathbf{o})$ onto the spherical harmonics $B_k(\mathbf{o})$ according to the outgoing direction, obtaining the matrix $c^t_{jk}$, which corresponds to $B_j(\boldsymbol{\omega})$ and $B_k(\mathbf{o})$ respectively as follows:

\begin{equation}%
    \label{eqn:view_dependent_4}
    c^t_{jk} =  \int{B_k(\mathbf{o})} c_j^t(\mathbf{o})d\mathbf{o}
\end{equation}%

By using the coefficients $c^t_{jk}$ and the corresponding basis functions $B_k(\mathbf{o})$, it is also possible to approximate $c^t_j(\mathbf{o})$ as shown below.

\begin{equation}%
    \label{eqn:view_dependent_5}
    c^t_{j}(\mathbf{o}) \approx  \sum_{k=1}^{n^2} B_k(\mathbf{o}) c^t_{jk}
\end{equation}%

Substituting Equation \ref{eqn:view_dependent_5} into Equation \ref{eqn:view_dependent_3}, we can obtain an approximation of the outgoint radiance as follows:

\begin{equation}%
    \label{eqn:view_dependent_6}
{L_s}(\mathbf{o}) \approx \boldsymbol{\rho}_s \sum_{j=1}^{n^2} \sum_{k=1}^{n^2} {c_j}c^t_{jk}B_k(\mathbf{o})
\end{equation}%

However, computing the radiance transfer matrix, which stores the coefficients $c^t_{jk}$, is required for modeling view-dependent effects. For $n$-th order spherical harmonics (SH) lighting, each transfer matrix requires $n^2$ parameters. Consequently, with increasing numbers of Gaussians, the storage cost grows rapidly and becomes impractical. To address this, our final computation adopts the formulation in Equ.~\ref{eqn:view_dependent_3}, where $c^t_j(\mathbf{o})$ is dynamically decoded from the radiance transfer features $f_t$  and the reflection direction $\mathbf{o}$ by a lightweight MLP $G$ as follows:
\begin{equation}%
    \label{eqn:radiance_transfer_specular}
{L_s(\mathbf{o})} \approx \boldsymbol{\rho_s}\sum_{j=0  }^{n^2}{c_j} c^t_j(\mathbf{o}), \quad with \quad c^t_j(\mathbf{o}) = G(f_t,\mathbf{o})
\end{equation}%

This design significantly alleviates the storage overhead while maintaining flexibility in modeling view-dependent appearance.

\section{BRDF MODEL}
We adopt the microfacet specular shading model according to:

\begin{equation}%
    \label{eqn:6}
        f(\mathbf{i}, \mathbf{o}) = \frac{DFG}{4(\mathbf{n} \cdot \mathbf{o})(\mathbf{n} \cdot \mathbf{i})}
\end{equation}%

\noindent where $D, F, G$ correspond to normal distribution function, fresnel term, and geometry term. Their specific expressions
are as follows:

\begin{equation}%
    \label{eqn:7}
        D(\mathbf{n}, \mathbf{h}, a) = 
        \frac{a^2}{\pi((\mathbf{n} \cdot \mathbf{h})^2 (a^2-1) +1)^2}
\end{equation}%

\begin{equation}%
    \label{eqn:8}
        F = F_0 + (1-F_0) (1- (\mathbf{h} \cdot \mathbf{o}))^5
\end{equation}%

\begin{equation}%
    \label{eqn:9}
        G(\mathbf{n}, \mathbf{o}, \mathbf{i}, k) 
        = G_{sub}(\mathbf{n}, \mathbf{o}, k) \cdot G_{sub}(\mathbf{n}, \mathbf{i}, k) 
\end{equation}%

\begin{equation}%
    \label{eqn:10}
    G_{sub}(\mathbf{n}, \mathbf{v}, k) 
    = \frac{\mathbf{n} \cdot \mathbf{v}}{(\mathbf{n} \cdot \mathbf{v}) (1-k) + k}
\end{equation}%

\noindent where $\mathbf{n}$ is normal, $\mathbf{h}$ is half-way vector. Roughness $r$ determines $a$ and $k$, where $a = r^2$ and $k = \frac{r^4}{2}$. $F_0$ in $F$ is the basic reflection ratio, calculated by metallic $m$ and albedo $\mathbf{c}$ as follows:

\begin{equation}%
    \label{eqn:11}
    F_0 = (1-m) * 0.04 + m * \mathbf{c}
\end{equation}%

\section{IMPLEMENTATION DETAILS}

We conducted comprehensive experiments using an NVIDIA RTX 4090 GPU and used the Adam optimizer for all parameter updates. For object level data, Our hybrid rendering branch rendering speed is 96.4 and PBR branch is 130.9 in FPS, exhibiting the real-time rendering capability of our proposed inverse rendering method.

The model is first trained for 30,000 iterations using only the hybrid rendering branch. The view-independent components of radiance transfer are initialized at the beginning of training. After 3,000 iterations, the view-dependent components are activated. During initialization, the reflection intensity is set to 0.01 for all Gaussians, and the radiance transfer order is set to 3.
The MLP $G$ consists of three layers with 64 units each. It takes the reflection direction as input and concatenates transfer features at the second layer. The first two layers use the ReLU activation function. Additionally, a ReLU operation is applied after the dot product with the spherical harmonic lighting.
After training for 30,000 iterations, visibility information is baked into voxel grids with a resolution of $128^3$. Subsequently, both the hybrid rendering and physically-based rendering branches are jointly supervised for another 10,000 iterations. During this stage, the geometry is fine-tuned, and the appearance is decomposed into material and lighting components.
Both the reflection map and environment map are configured with a resolution of $6 \times 128 \times 128$, which balances computational efficiency and rendering quality.

In addition to the optimization mentioned in the main text, we also include the following commonly used loss terms:

\noindent\textbf{Bilateral Smoothness.} We believe that normal $\mathbf{n}$, reflection intensity $R_i$, reflection roughness $R_r$, metallic $m$, and roughness $r$ will not change drastically in color-smooth regions. We define a smooth constraint as:

\begin{equation}%
    \label{eqn:Bilateral Smoothness}
    \mathcal{L}_{s,f}=\Vert \nabla f \Vert 
    exp(-\Vert \nabla C_{gt} \Vert)
\end{equation}%

\noindent where $f$ represents the screen-space buffer of above attributes. For each term, the corresponding $\lambda_f = 0.01$.

\begin{equation}%
    \label{eqn:Bilateral Smoothness}
    \mathcal{L}_{s}= \sum\lambda_f \mathcal{L}_{s,f}
\end{equation}%

\noindent\textbf{Object Mask Constraint}. If there is a mask indicating the object, we can
constrain the optimization by a binary cross-entropy loss:

\begin{equation}%
    \label{eqn:Object Mask Constraint}
    \mathcal{L}_o = -M \log O - (1-M) \log (1-O)
\end{equation}%

\noindent where $M$ is the mask of the object and $O = \sum_i^N T_i\alpha_i$, and the corresponding  $\lambda_o = 0.1$

\section{MORE COMPARISONS}

We provide additional results for relighting and novel view synthesis to enable a comprehensive comparison. Notably, the ball from the Shiny Blender dataset does not include ground truth (GT) relighting data. However, we showcase our results of the ball to highlight the performance advantages of our method.

\begin{table}[]
  \caption{Albedo decomposition quality comparison.}
  \label{tab:Albedo decomposition}
  \begin{tabular}{cc|cccc}
  \hline
  & & TensoIR & GS-IR & R3DG & Ours \\
  \hline
    \multirow{2}{*}{PSNR↑}
    &TensoIR        & 29.19 & \textbf{32.04} & 28.27 & \underline{31.97} \\
    &Shiny Blender   & \underline{22.17} & 20.97 & 20.69 & \textbf{24.47} \\
  \hline
    \multirow{2}{*}{SSIM↑}
    &TensoIR        & \textbf{0.952} & 0.920 & 0.918 & \underline{0.939} \\
    &Shiny Blender   & \underline{0.877} & 0.859 & 0.871 & \textbf{0.913} \\
  \hline
    \multirow{2}{*}{LPIPS↓}
    &TensoIR        & 0.080 & 0.092 & \underline{0.070} & \textbf{0.052}\\
    &Shiny Blender  & 0.184 & 0.160 & \underline{0.141} & \textbf{0.085}\\
  \hline
  \end{tabular}
\end{table}

We present quantitative evaluations of albedo decomposition using PSNR, SSIM, and LPIPS on both the Shiny Blender dataset and the TensoIR dataset, as shown in Table~\ref{tab:Albedo decomposition}.

\begin{figure}[]
  \includegraphics[width=\linewidth]{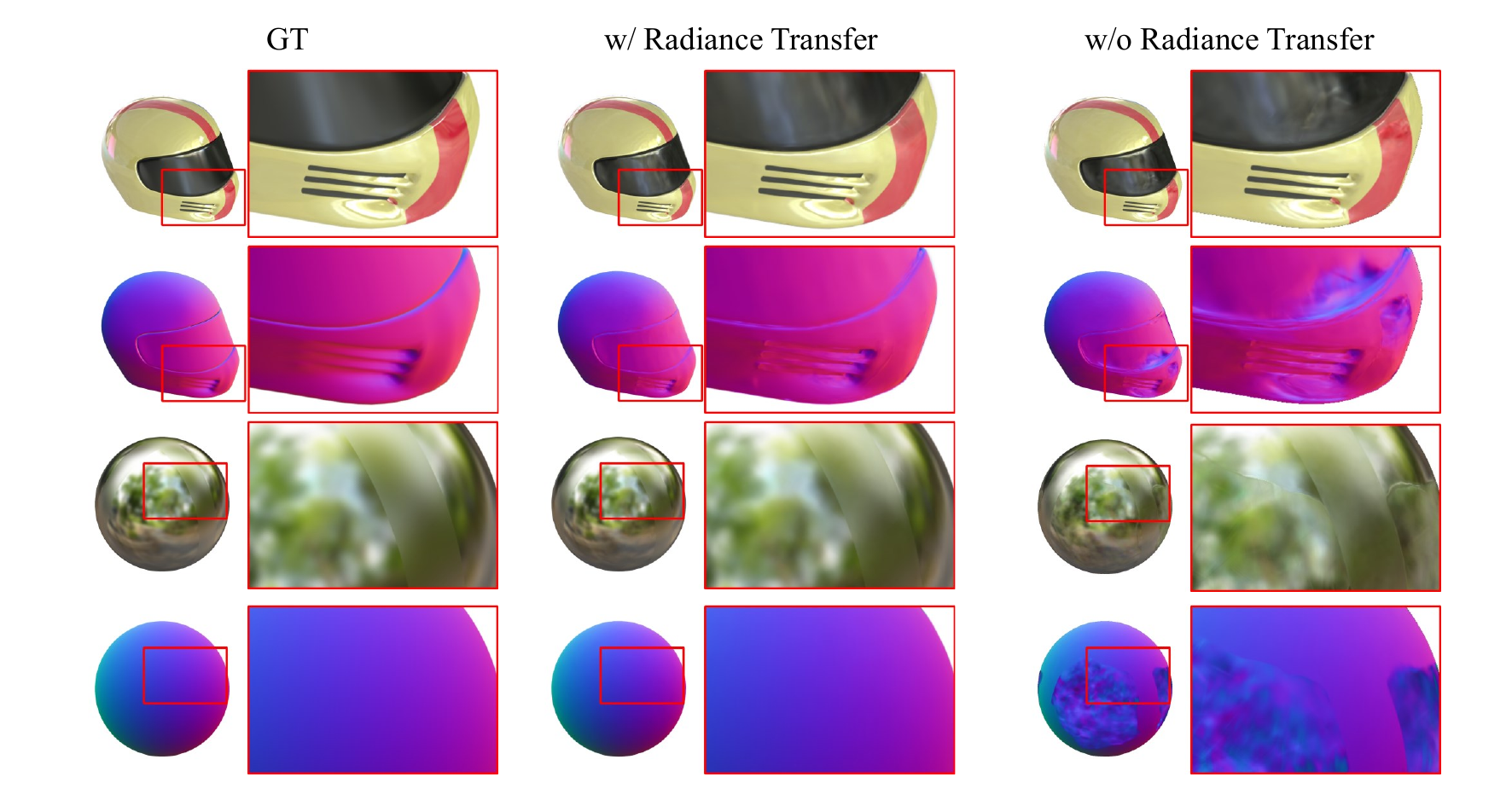}
  \caption{Visual comparison of using radiance transfer and spherical harmonics}
  \label{fig:without-prt}
\end{figure}

Figure~\ref{fig:without-prt} visualizes the radiance transfer component of our model. This component effectively enhances surface normals. Our radiance transfer maintains low-frequency characteristics better than spherical harmonics, preventing the appearance of floating artifacts and resulting in smoother surfaces with more accurate normals.

\begin{figure}[]
  \includegraphics[width=\linewidth]{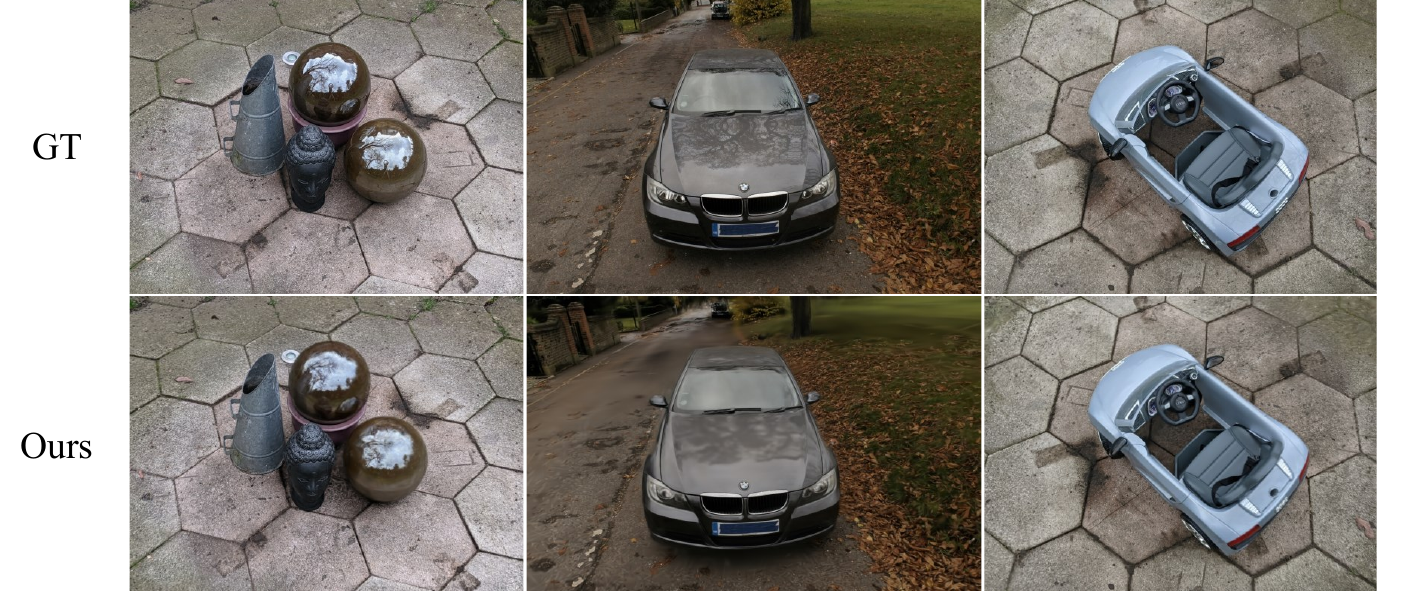}
  \caption{Qualitative comparisons on real scenes.}
  \label{fig:ref-real}
\end{figure}

\begin{figure}[]
  \includegraphics[width=\linewidth]{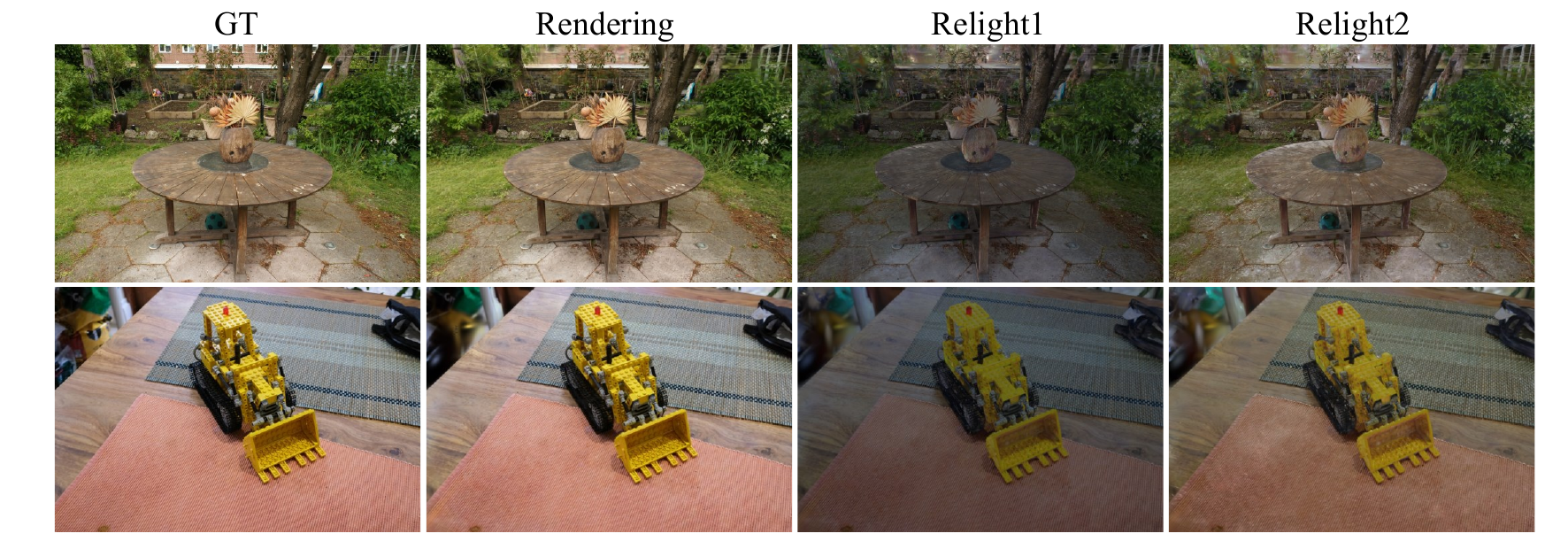}
  \caption{Relighting results on real scene dataset.}
  \label{fig:mipnerf360-relight}
\end{figure}

Figure~\ref{fig:ref-real} shows the results on the Ref-Real dataset, where our method achieves high-quality performance even on real-world data without requiring masks. Additionally, we performed relighting tests on the kitchen and garden scenes from the Mip-NeRF 360 dataset, as shown in Figure~\ref{fig:mipnerf360-relight}.

\begin{figure*}[tbp]
  \includegraphics[width=\textwidth]{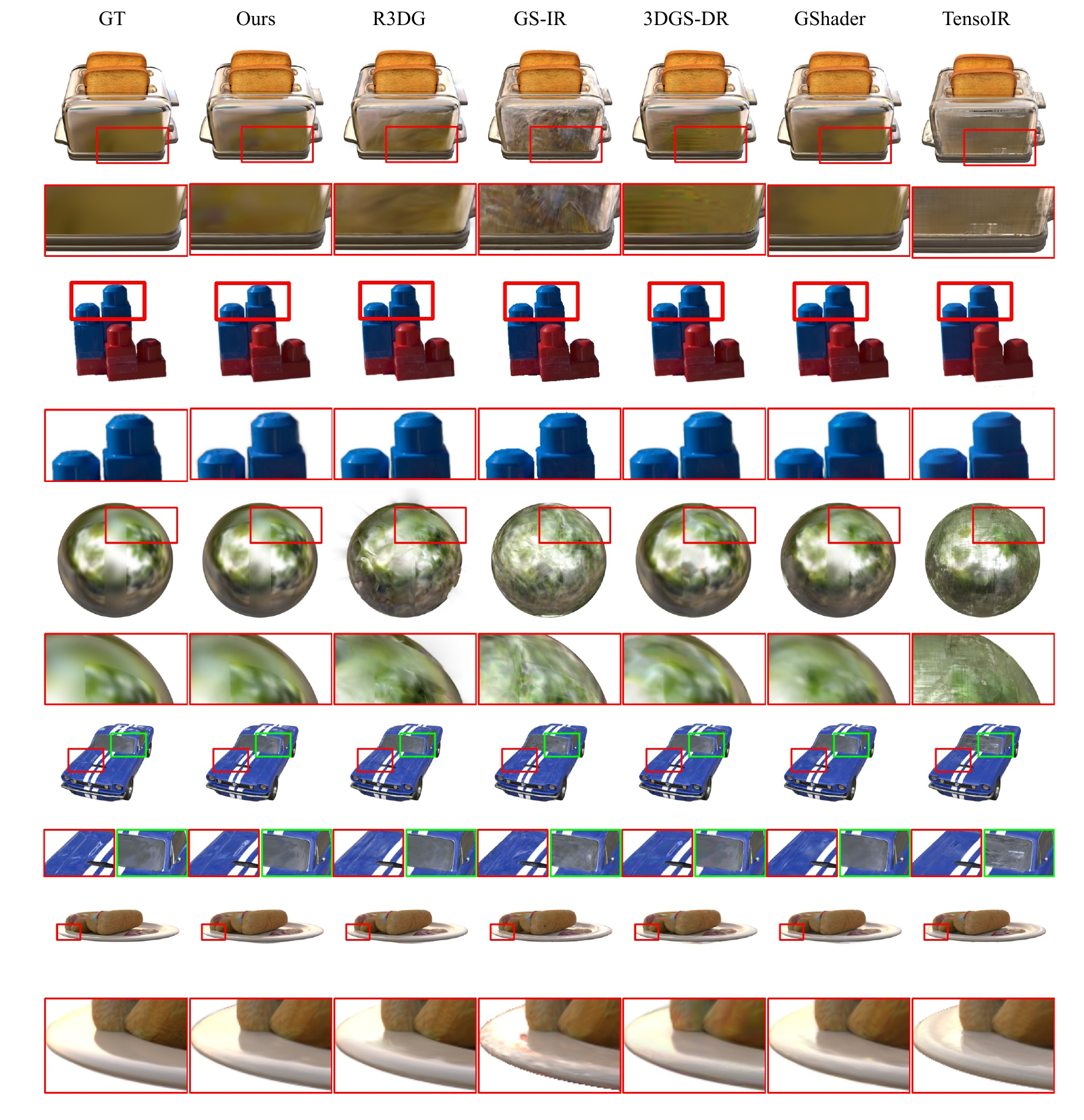}
  \caption{Qualitative comparisons on synthetic scenes.}
  \label{fig:sup-nvs}
\end{figure*}

Figure~\ref{fig:sup-nvs} compares the novel view synthesis results of our method (RTR-GS) with those of other approaches. Enlarged views of local regions are included to emphasize details. Our method demonstrates superior reflection clarity and captures finer details compared to others.

\begin{figure*}[tbp]
  \includegraphics[width=\textwidth]{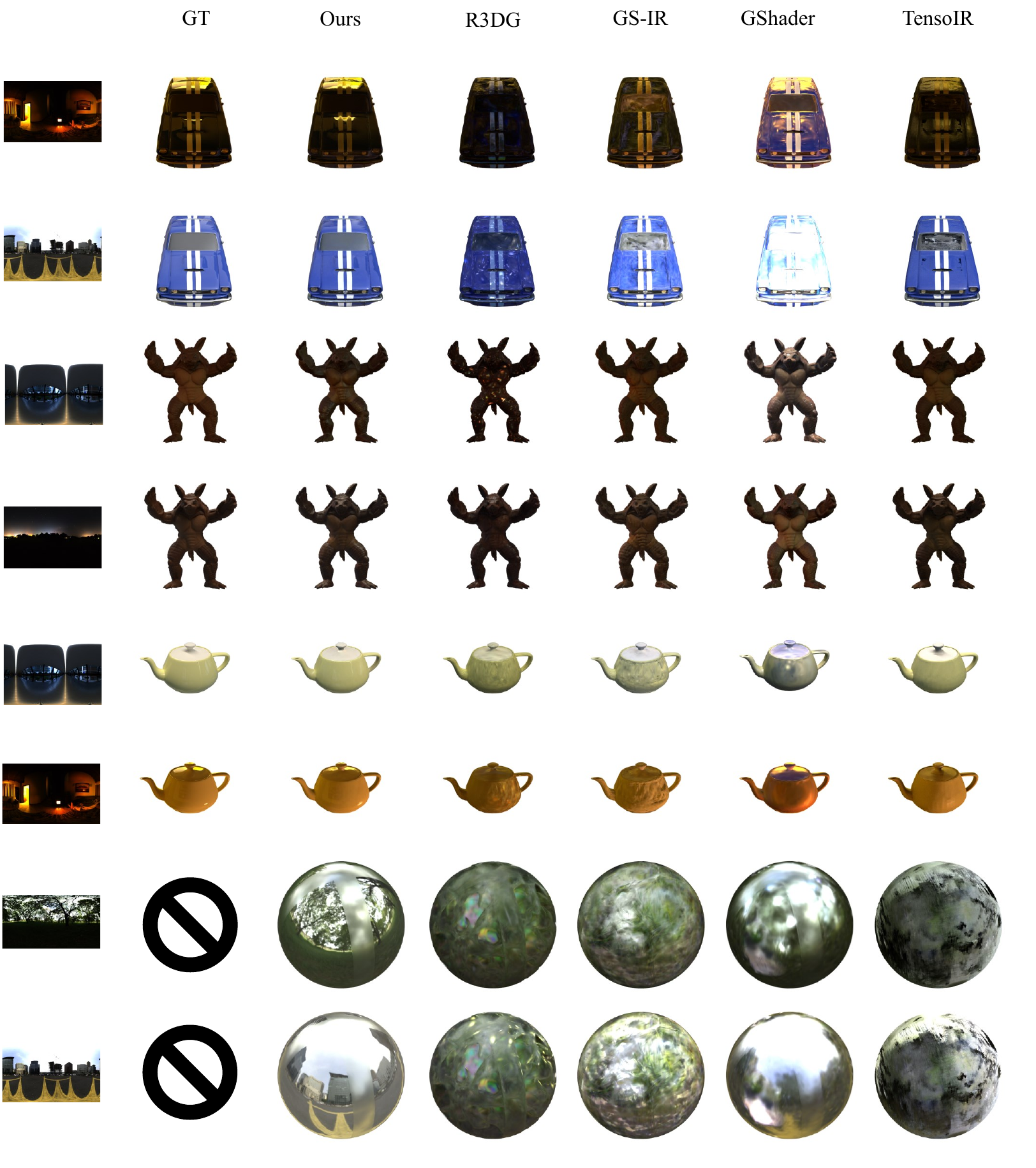}
  \caption{Qualitative comparisons of relighting with different environment lighting.}
  \label{fig:sup-relight-comp}
\end{figure*}

Figure~\ref{fig:sup-relight-comp} shows a comparison of relighting results between our approach and other inverse rendering methods. For both diffuse and specular objects, our method produces more accurate and realistic relighting outcomes. Specifically, reflective surfaces exhibit precise and detailed reconstructions of reflections. Additionally, as shown in Table~\ref{tab: sub-relight-objects}, we provide quantitative relighting results for selected objects from the datasets used.

\begin{figure}[]
    \includegraphics[width=\linewidth]{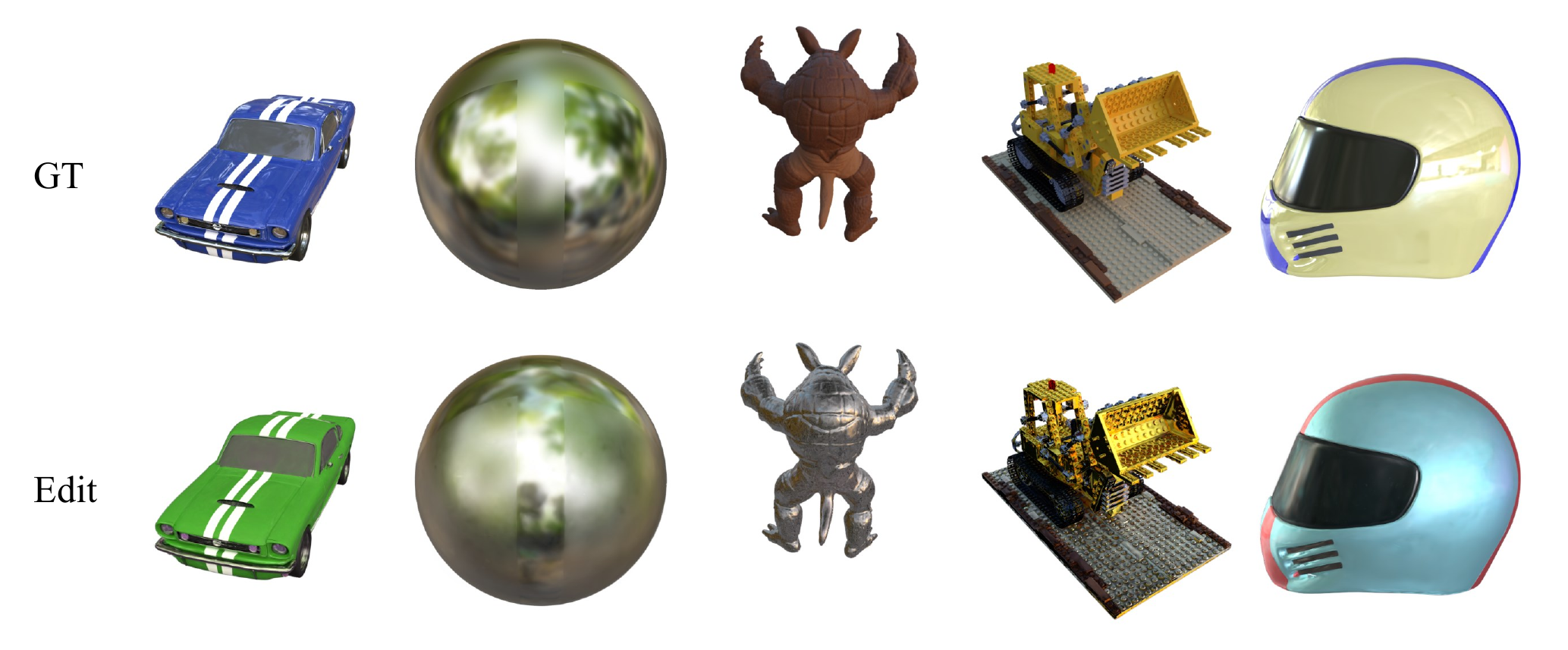}
  \caption{Material editing results.}
  \label{fig:materials-editing}
\end{figure}

\begin{figure*}[tbp]
  \includegraphics[width=\textwidth]{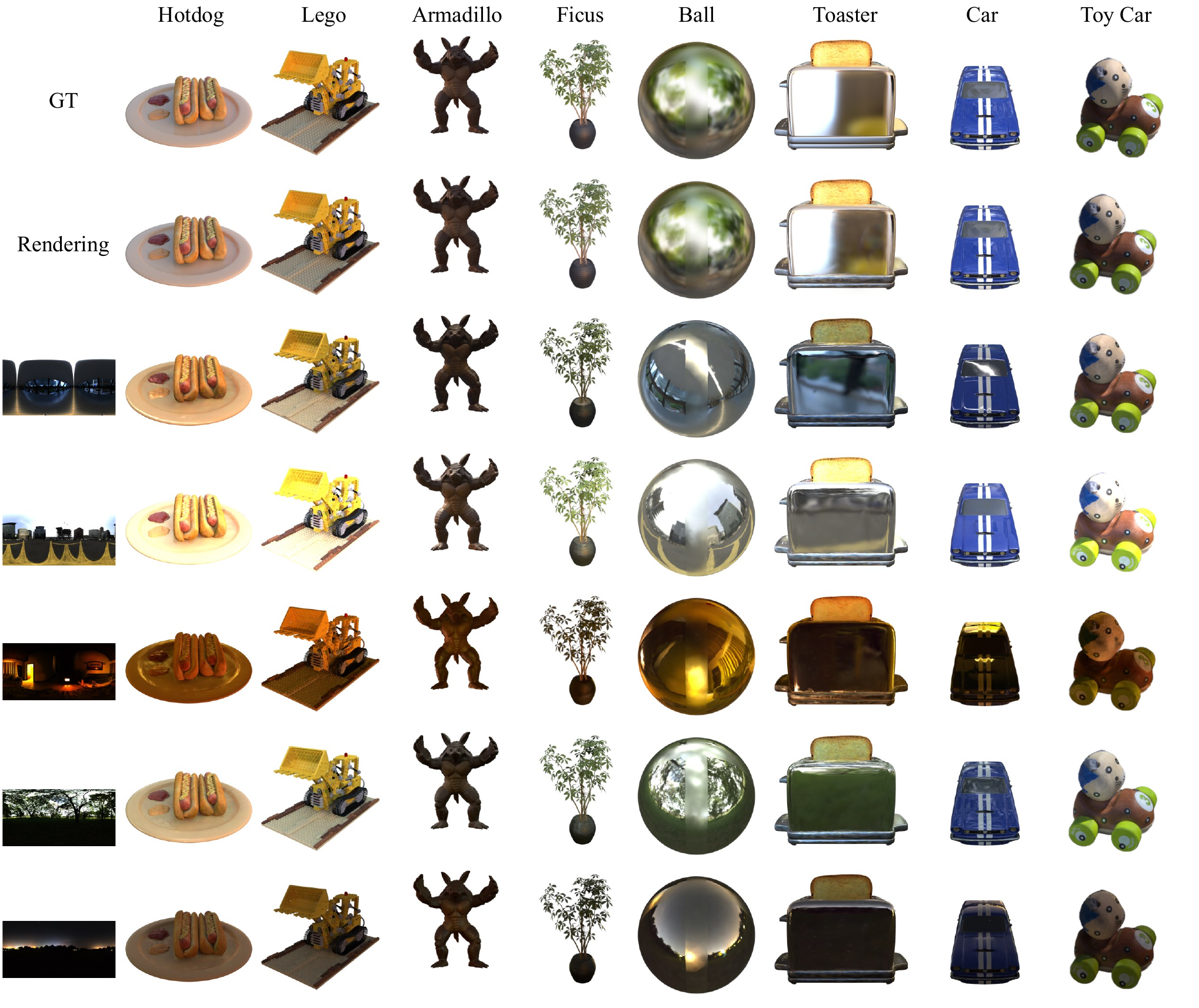}
  \caption{Relighting results of our method on synthetic dataset. Our method can also provide high-quality relighting results for diffuse
objects and specular objects.}
  \label{fig:sup-relight}
\end{figure*}

Figure~\ref{fig:sup-relight} presents the results generated by our method under five different lighting conditions. The outputs are consistent and accurate for both diffuse and specular objects, demonstrating the robustness of our approach. Furthermore, as shown in Figure~\ref{fig:materials-editing}, our method also produces reliable results after material editing.

\begin{table}[]
  \caption{Relighting quality of some objects in terms of PSNR↑ and SSIM↑ on the TensoIR dataset (top 4 rows), Shiny Blender dataset (middle 5 rows), and Stanford ORB dataset (bottom 5 rows).}
  \label{tab: sub-relight-objects}
  \begin{tabular}{ccccc}
  \hline
  & TensoIR & R3DG & GS-IR & Ours \\
  \hline
   hotdog & \underline{27.87}/\underline{.932} & 24.40/.922 & 26.86/.921 & \textbf{28.91}/\textbf{.943} \\
   ficus & 24.30/\underline{.946} & \underline{28.74}/.941 & 23.08/.872 & \textbf{31.05}/\textbf{.953} \\
   lego  & \underline{27.57}/\textbf{.924} & \textbf{28.23}/\textbf{.924} & 21.27/.854 & 25.68/\underline{.914} \\
   armadillo & \underline{34.46}/\textbf{.975} & 32.70/.951 & 32.73/.941 & \textbf{34.76}/\underline{.967} \\
  \hline
   car      & \underline{26.15}/\underline{.913} & 22.37/.881 & 23.58/.872 & \textbf{28.74}/\textbf{.947} \\
   coffee   & 18.37/.845 & 16.76/.865 & \underline{19.95}/\underline{.877} & \textbf{19.98}/\textbf{.909} \\
   helmet   & 17.28/.791 & \underline{19.31}/\underline{.848} & 17.99/.797 & \textbf{24.44}/\textbf{.918} \\
   teapot   & \underline{33.73}/\underline{.979} & 27.44/.976 & 29.59/.969 & \textbf{35.68}/\textbf{.989} \\
   toaster  & 15.95/.682 & \underline{17.54}/\underline{.774} & 14.80/.715 & \textbf{21.97}/\textbf{.879} \\
  \hline
  baking\_001 & \underline{26.53}/.961 & \textbf{26.70}/\textbf{.969} & 25.95/\underline{.965} & 25.71/\textbf{.969} \\
  car\_002    & 26.65/.964 & 28.95/.963 & \underline{29.54}/\underline{.965} & \textbf{29.69}/\textbf{.975} \\
  chips\_002  & 28.65/.947 & 32.32/.969 & \underline{33.46}/\underline{.973} & \textbf{33.71}/\textbf{.974} \\
  grogu\_001  & 25.73/.959 & 27.37/.968 & \textbf{28.55}/\underline{.970} & \underline{27.39}/\textbf{.972} \\
  \hline
  \end{tabular}
\end{table}

\end{document}